\documentclass[10pt]{article}

\usepackage{fancyhdr}
\usepackage{extramarks}
\usepackage{amsmath}
\usepackage{amsthm}
\usepackage{amsfonts}
\usepackage{siunitx}
\usepackage{tikz}
\usepackage[plain]{algorithm}
\usepackage{algpseudocode}
\usepackage{multirow}
\usepackage{booktabs}
\usepackage{palatino}
\usepackage{graphicx}
\usepackage{subfigure}
\usepackage[colorlinks,linkcolor=black,anchorcolor=black,citecolor=black,urlcolor=blue]{hyperref}
\usepackage{amsmath,bm}
\usepackage{booktabs}
\usepackage{mathtools}
\usepackage{amssymb}
\usepackage{caption}
\usepackage{capt-of}
\usepackage{mciteplus}
\usepackage{cite}
\usepackage{mathrsfs}
\usepackage[title,titletoc,toc]{appendix}
\usepackage{xr}
\usepackage{parskip}
\usetikzlibrary{automata,positioning}

\renewcommand{\part}[1]{\textbf{\large Part \Alph{partCounter}}\stepcounter{partCounter}\\}

\begin{document}

\title{Decoding SARS-CoV-2 transmission, evolution and ramification on COVID-19 diagnosis, vaccine, and  medicine}
\author{ Rui Wang$^1$, Yuta Hozumi$^1$,  Changchuan Yin$^2$ \footnote{Address correspondences to Changchuan Yin. E-mail:cyin1@uic.edu}, and Guo-Wei Wei$^{1,3,4}$ \footnote{Address correspondences to Guo-Wei Wei. E-mail:wei@math.msu.edu} \\
$^1$ Department of Mathematics,
Michigan State University, MI 48824, USA\\
$^2$ Department of Mathematics, Statistics, and Computer Science, \\
University of Illinois at Chicago, Chicago, IL 60607, USA\\
$^3$  Department of Biochemistry and Molecular Biology\\
Michigan State University, MI 48824, USA \\
$^4$ Department of Electrical and Computer Engineering \\
Michigan State University, MI 48824, USA }

\date{}

\maketitle

\begin{abstract}
 Tremendous effort has been given to the development of diagnostic tests, preventive vaccines, and therapeutic medicines for coronavirus disease 2019 (COVID-19) caused by severe acute respiratory syndrome coronavirus 2 (SARS-CoV-2). Much of this development has been based on the reference genome collected on January 5, 2020. Based on the genotyping of 6156 genome samples collected up to April 24, 2020, we report that SARS-CoV-2 has  had 4459 alarmingly  mutations which can be clustered into five subtypes.   We introduce mutation ratio and mutation $h$-index to characterize the protein conservativeness and unveil that  SARS-CoV-2 envelope protein, main protease, and endoribonuclease protein are relatively conservative, while SARS-CoV-2 nucleocapsid protein, spike protein, and papain-like protease are relatively non-conservative. In particular, the nucleocapsid protein has more than half its genes changed in the past few months, signaling  devastating impacts on the ongoing development of COVID-19 diagnosis, vaccines, and drugs.

\end{abstract}


\newpage

\renewcommand{\thepage}{{\arabic{page}}}

\section{Introduction} 
The ongoing pandemic of coronavirus disease 2019 (COVID-19) caused by severe acute respiratory syndrome coronavirus 2 (SARS-CoV-2) poses crucial threats to the public health and the world economy since it was detected in Wuhan, China in December 2019 \cite{gorbalenya2020species}. As of April 24, 2020, more than 2.6 million cases of COVID-19 have been reported in 185 countries and territories, resulting in more than 184,000 deaths \cite{who_2020}. Tragically, there is no sign of slowing down nor relief at this monument partially due to the fact there is no specific anti-SARS-CoV-2 drugs and effective vaccines. 

SARS-CoV-2 is a positive-strand RNA virus and belongs to the beta coronavirus genus. The genomic information underpins the development of   antiviral medical interventions, preventative vaccines, and viral diagnostic tests. The first SARS-CoV-2 genome was reported on January 5, 2020 \cite{wu2020new}. It has a genome size of 29.99 kb, which encodes for multiple non-structural and structural proteins. The leader sequence and ORF1ab encode non-structural proteins for RNA replication and transcription. Among various non-structural proteins, viral  papain-like (PL) proteinase, main protease (or 3CL protease), RNA polymerase, and endoribo-nuclease are common targets in antiviral drug discovery. However, it typically takes more than ten years to put an average drug to the market. The downstream regions of the genome encode structural proteins including spike (S) protein, envelope (E) protein, membrane (M) protein, and nucleocapsid (N) protein. Notably, S-protein uses one of two subunits to bind directly to the host receptor angiotensin-converting enzyme 2 (ACE2), enabling the virus entry into host cells \cite{xiao2003sars}. The nucleocapsid (N) protein, one of the most abundant viral proteins, can bind to the RNA genome and is involved in replication processes, assembly, the host cellular response during viral infection \cite{mcbride2014coronavirus}. The E protein is a small integral membrane protein, a virulence factor, regulating cell stress response and apoptosis, and promoting inflammation \cite{dediego2011severe}. The structural proteins, especially, the Spike protein and the N protein, are the candidate antigens for vaccine development. Developing safe and effective vaccines is urgently needed to prevent the spread of SARS-CoV-2. However, it typically takes over one year to design and test a new vaccine. Furthermore, the SARS-CoV-2 genome undergoes rapid mutations partially stimulated as a response to the challenging immunological environments arising from the COVOD-19 patients of different races, ages, and medical conditions. SARS-CoV-2 exists as heterogeneous and dynamic populations because of their error-prone replication \cite{moya2004population}. 

 The vaccine developed at one time may not be effective for mitigating the infection by new mutated virus isolates. An alarming fact is that many of these mutations may devastate the on-going effort in the development of effective medicines,   preventive vaccines, and diagnostic tests. Accurate identification of the antigens and their mutations represents the most important roadblock in developing effective vaccines against COVID-19. For example, different vaccines are needed for different geographic locations due to predominant mutations in the corresponding regions. In COVID-19 diagnosis, the diagnosis kits are designed using two major methods, i.e., specific serological tests and molecular tests. Serological tests are to detect specific COVID-19 proteins. Molecular diagnoses test specific COVID-19 pathogenic genes, which usually rely on the polymerase chain reaction (PCR).Because of the fast mutations of the SARS-CoV-2 genome, genotyping analysis of SARS-CoV-2 may optimize the PCR primer design to detect SARS-CoV safely and reduce the risk of false-negatives caused by genome sequence variations. In addition, the genotyping analysis may also reveal those regions that are highly conserved with very few mutations, which can be selected as a target sequence for reliable drug therapy and general diagnosis.

The evolution pattern through the highly frequent mutations of SARS-CoV-2 can be observable on short time scales. In the early infection period (i.e., February 2020), the SARS-CoV-2 variants were clustered as S and L types \cite{zhang2020origin}. Recent genotyping analysis reveals a large number of mutations in various essential genes encoding the S protein, the N protein, and RNA polymerase in the SARS-CoV-2 population \cite{yin2020genotyping}. Monitoring the evolutionary patterns and spread dynamics of SARS-CoV-2 is of grace importance for COVID-19 control and prevention.

Although mutations occur randomly, most preserved mutations can be regarded as  virus responds to the host immune system surveillance. As a result, the faster and the wider the SARS-CoV-2 spread is, the more frequent and diverse the mutations will be. The tracking and analysis of COVID-19 dynamics, transmission, and spread is of paramount importance for winning the on-going battle against COVID-19. Genetic identification and characterization of the geographic distribution, intercontinental evolution, and global trends of SARS-CoV-2 is the most efficient approach for studying COVID-19 genomic epidemiology and offer the molecular foundation for region-specific SARS-CoV-2 vaccine design, drug discovery, and diagnostic development \cite{ojosnegros2010models}. For example, different vaccines for the shell can be designed according to predominant mutations.

This work provides the most comprehensive genotyping to reveal the transmission trajectory and spread dynamics of COVID-19 to date. Based on genotyping 6156 SARS-CoV-2 genomes from the world as of April 24, 2020, we trace the COVID-19 transmission pathways and analyze the distribution of the subtypes of SARS-CoV-2 across the world. We use $K$-means methods to cluster SARS-CoV-2 mutations, which provides the updated molecular information for the region-specific design of vaccines, drugs, and diagnoses. Our clustering results show that globally, there are at least five distinct subtypes of SARS-CoV-2 genomes. While, in the U.S., there are at least three significant SARS-CoV-2 genotypes. We introduce mutation $h$-index and mutation ratio to characterize conservative and non-conservative proteins and genes.  We unveil the unexpected non-conservative genes and proteins, rendering an alarming warning for the current development of diagnostic tests, preventive vaccines, and therapeutic medicines.  

\section{Results and discussion}

\subsection{COVID-19 evolution and clustering}

Tracking the SARS-CoV-2 transmission pathways and analyzing the spread dynamics are critical to the study of genomic epidemiology. Temporospatially clustering the genotypes of SARS-CoV-2 in transmission  provides insights into diagnostic testing and vaccine development in disease control. In this work, we retrieve and genotype 6156 SARS-CoV-2 isolates from   word as of April 24, 2020. There are 4459 single mutations in  6156 SARS-CoV-2 isolates.  Based on these mutations, we classify and track the geographical distributions of 6156 genoytype isolates by $K$-means clustering. The SARS-CoV-2 genotypes, represented as SNP variants, are clustered as five groups in the world \autoref{table:World}. The genotypes in the U.S. are clustered as three groups \autoref{table:World}. The optimal clustering groups   are established using the Elbow method in the $K$-means clustering algorithm (see Supporting Material). 

\begin{table}[H]
	\centering
	\setlength\tabcolsep{5pt}
	\captionsetup{margin=0.9cm}
	\caption{Co-mutations with the highest frequency in five distinct clusters.}
	\begin{tabular}{clc}
		\hline
	    Cluster  & Mutation sites  & Number of descendants      \\ \hline
		I  & [8782C$>$T, 28144T$>$C]  & 943 \\  
	    II  & [14408C$>$T] & 3857\\ 
	   	III  & [3037C$>$T, 14408C$>$T, 23403A$>$G] & 3842 \\
    	IV & [3037C$>$T, 14408C$>$T, 23403A$>$G, 28881G$>$A, 28882G$>$A, 28883G$>$C] &965 \\
		V & [241C$>$T, 3037C$>$T, 14408C$>$T, 23403A$>$G, 25563G$>$T]  & 3032 \\ \hline
	\end{tabular}
	\label{table:Frequency World}
\end{table}

\begin{figure}[ht]
    \centering
    \includegraphics[scale=0.5]{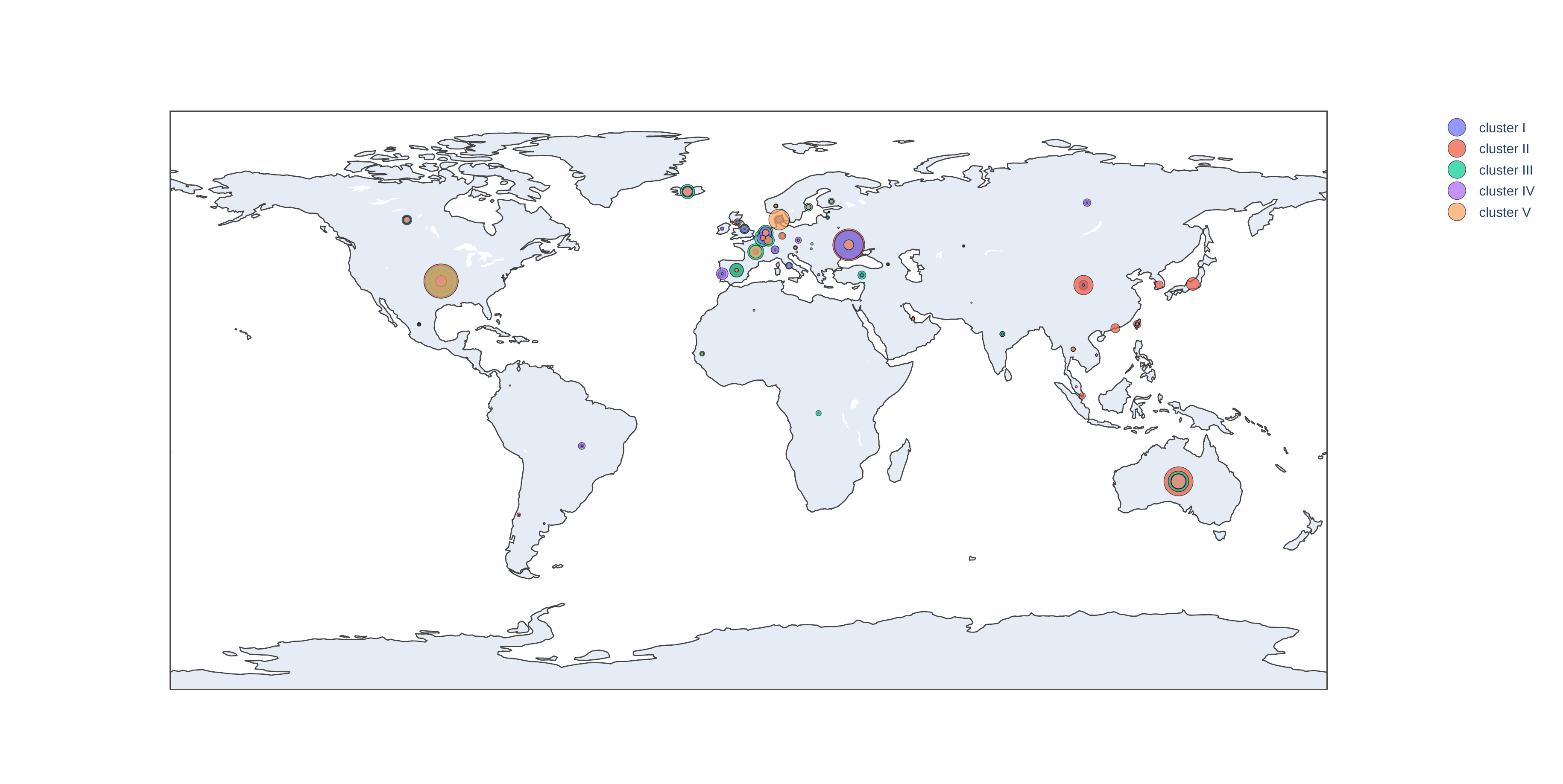}
    \caption{The scatter plot of five distinct clusters in the world. The blue, red, green, purple, and yellow circles represent clusters I - V. The size of each circle shows the number of SNP variants in each cluster.}
    \label{fig:World}
\end{figure}

The detailed distribution of the SNP variants from the world for each cluster is provided in the Supporting Material. The SNP variant clusters from 11 countries that have the highest number of cases recorded in are listed in \autoref{table:World}. The countries are United States (US), Canada (CA), Australia (AU), United Kingdom (UK), Germany (DE), France (FR), Italy (IT), Ukraine (UA), China (CN), Japan (JP), and Korean (KR). The numbers of SNP variants in each cluster are shown on the world map in \autoref{fig:World}. The geographic distribution of the SNP variant clusters reflects the approximate transmission pathways and spread dynamics across the world. Several findings can be read from the \autoref{table:World}:
\begin{enumerate}
    \item[1.] Two early subtypes of SASR-CoV-2 (cluster I and II) are epidemic in the Asian countries (CN, JP, KR).
    \item[2.] The subtypes of SARS-CoV-2 in Cluster III is not spreading in the European countries (UK, DE, FR, IT). 
    \item[3.] All of the subtypes of SARS-CoV-2 in five different clusters can be found in the US, AU, and Canada. 
    \item[4.] The dominant subtypes of SARS-CoV-2 epidemic in the United States belong to all the five clusters.
\end{enumerate}
 The cluster analysis reveals that the Asian countries only have two dominant subtype clusters, cluster I [8782C$>$T, 28144T$>$C], and cluster II [14408C$>$T]. Cluster I was detected in the early period of COVID-19 infection in China and Asian countries. The subtype of SNP mutation in Spike protein, 23403A$>$G, is prevalent in the clusters III, IV, and V  of European countries. This subtype of Spike protein mutation may confer the high spread of SARS-CoV-2 in the European countries.

\begin{table}[H]
    \centering
    \setlength\tabcolsep{15pt}
    \captionsetup{margin=0.9cm}
    \caption{The cluster distributions of samples from  11 countries.}
    \begin{tabular}{lccccc}
    \hline
     Country        & Cluster I   & Cluster II  & Cluster III  & Cluster IV  & Cluster V  \\ \hline
     US             & 456        & 232        & 406         & 50         & 481   \\
     CA             & 40         & 13         & 28          & 14         & 10    \\
     AU             & 63         & 342        & 173         & 97         & 87   \\
     UK             & 0          & 23         & 10          & 4          & 0    \\
     DE             & 0          & 12         & 3           & 8          & 20    \\ 
     FR             & 0          & 14         & 105         & 6          & 66    \\
     IT             & 0          & 6          & 20          & 14         & 0     \\
     UA             & 2          & 414        & 322         & 348        & 41    \\
     CN             & 35         & 152        & 0           & 0          & 0    \\
     JP             & 0          & 67         & 0           & 0          & 0    \\
     KR             & 0          & 26         & 0           & 0          & 0    \\ \hline
    \end{tabular}
    \label{table:World}
\end{table}

\begin{figure}[ht]
    \centering
    \includegraphics[scale=0.5]{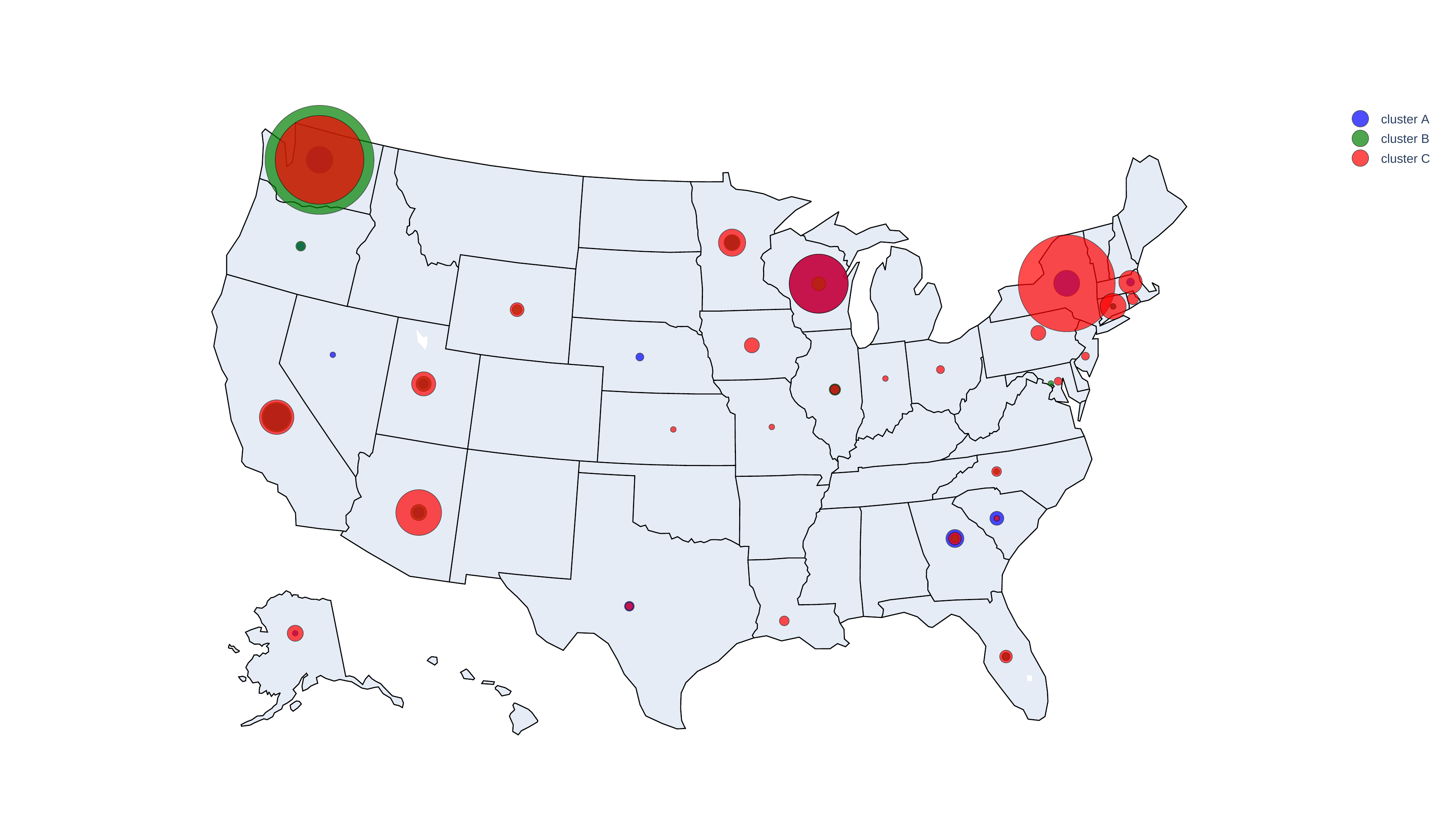}
    \caption{The scatter plot of three distinct clusters in the US. The blue, green, and red circles represent for cluster A, cluster B, and cluster C. The size of each circle reflects the number of SNP variants in each cluster.}
    \label{fig:US_Statistic}
\end{figure}

Moreover, we analyze the statistic of SNP variants located in different states of the United States. In \autoref{table:US}, we list the number of cases in three different clusters with respects to the west coast states (Washington (WA), California (CA), Alaska (AK), and Oregon (OR)), the east coast cities and states ( New York (NY), Pennsylvania (PA),  Florida (FL),  Massachusetts (MA), Maryland (MD), Virginia (VA)), Minnesota (MN), Georgia (GA), Utah (UT), Connecticut (CT), Arizona (AZ), and Idaho (ID). Notably, several findings on the genotypes of clusters in the US are as follows:
\begin{enumerate}
   \item[1.] Cluster B is dominated by the west coast states, especially the state of Washington. 
    \item[2.] Cluster C is dominated by the east coast states, especially in the state of New York, and in the midwest state of Wisconsin, and the Midwest state of Wisconsin. 
    \item[3.] The subtypes of SARS-CoV-2 in Cluster A are spread out through the United States. 
\end{enumerate}
\autoref{fig:US_Statistic} is the scatter plot of the three distinct clusters in the US. The size of each circle represents the number of SNP variants. The red circle represents Cluster C, which covers most of the states in the U.S. The red circle is the SNP variants in Cluster C, which mainly spreads in Idaho (ID).

We note that Cluster C in the U.S. is derived from Cluster III in the world, with an additional mutation at the leader sequence 241. The high spread in New York is consistent with the high transmission of SARS-CoV-2 in the European countries, where the subtype in Cluster III is predominated.

\begin{table}[H]
    \centering
    \setlength\tabcolsep{20pt}
    \captionsetup{margin=0.9cm}
    \caption{The cluster distributions of samples from  17 states and cities in the US.}
    \begin{tabular}{lccccc}
    \hline
     State          & Cluster A   & Cluster B   & Cluster C   \\ \hline
     WA             & 22         & 368        & 244 \\
     CA             & 22         & 27         & 37  \\
     AK             & 1          & 0          & 8   \\
     OR             & 2          & 3          & 0   \\
     NY             & 21         & 0          & 291 \\
     PA             & 0          & 0          & 7   \\
     FL             & 2          & 2          & 5   \\
     MA             & 2          & 2          & 17  \\
     MD             & 0          & 0          & 2   \\
     VA             & 3          & 9          & 28  \\
     WI             & 108        & 6          & 108 \\
     MN             & 6          & 8          & 23  \\
     GA             & 10         & 2          & 5   \\
     UT             & 3          & 7          & 18  \\
     CT             & 1          & 1          & 21  \\
     AZ             & 4          & 8          & 65  \\
     ID             & 1          & 0          & 21  \\
     IL             & 3          & 4          & 3   \\
     Others         & 31         & 7          & 36  \\ 
     Total          & 239        & 450        & 936 \\ \hline
    \end{tabular}
    \label{table:US}
\end{table}

\begin{table}[H]
    \centering
    \setlength\tabcolsep{20pt}
    \captionsetup{margin=0.9cm}
    \caption{The mutation sites with highest frequency in each clusters.}
    \begin{tabular}{lcl}
    \hline
                    & Frequency  & Mutation sites                 \\ \hline
     Cluster A      & 56         & [11083C$>$Y]                        \\
     Cluster B      & 455        & [17747C$>$T, 17858A$>$G, 28144T$>$C]          \\ 
     Cluster C      & 892        & [241C$>$T, 3037C$>$T, 14408C$>$T, 23403A$>$G]      \\ \hline
    \end{tabular}
    \label{table:Frequency US}
\end{table}

\begin{figure}[ht]
\centering
\includegraphics[scale=0.45]{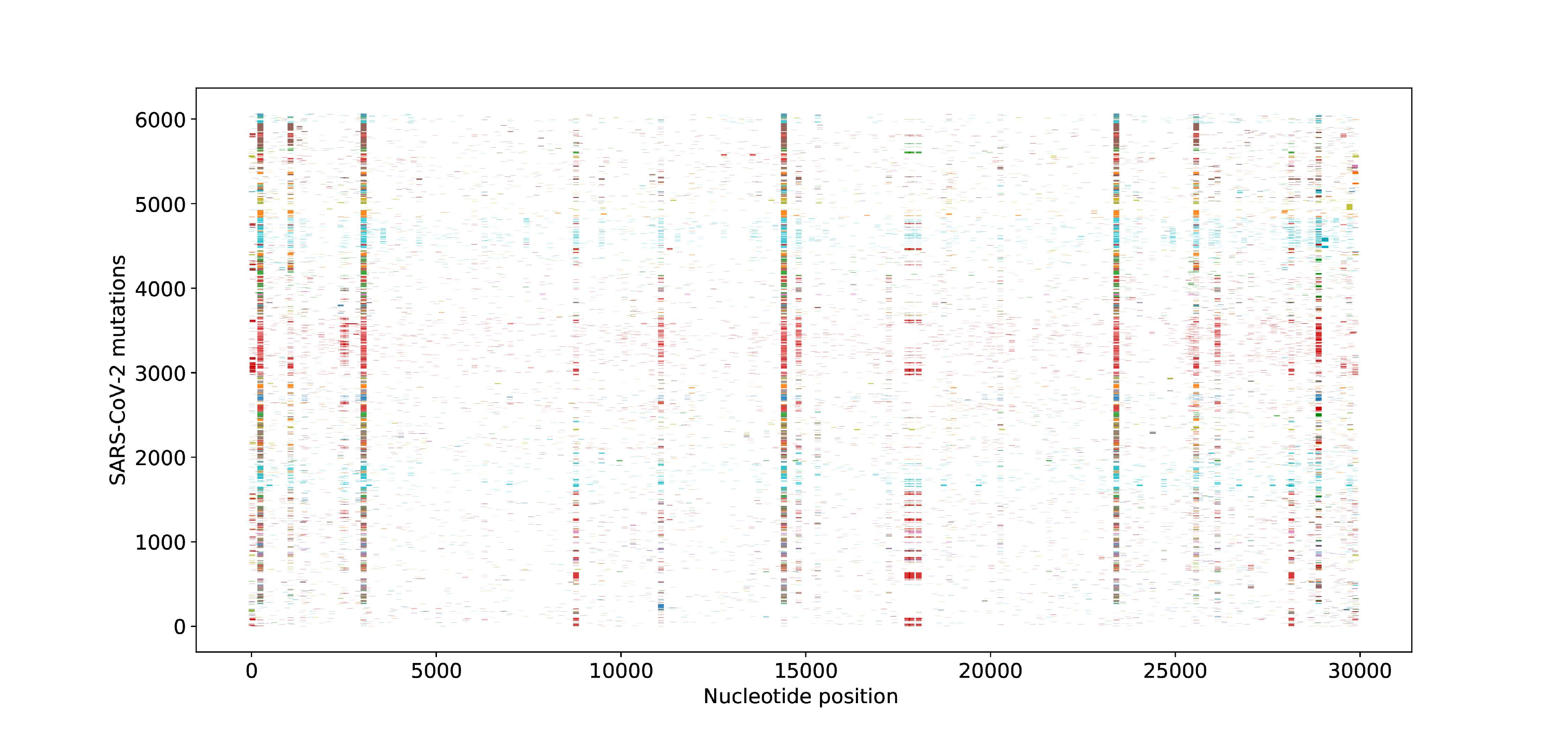}
\caption{Distribution of SNP mutations  of SARS-CoV-2 isolates from 6156 genome samples in the world with respect to the reference genome of January 5, 2020. (GenBank access number:  NC045512.2)}
\label{fig:subfig:Distribution}
\end{figure}

\begin{figure}[ht]
\centering
        \includegraphics[scale = 0.45]{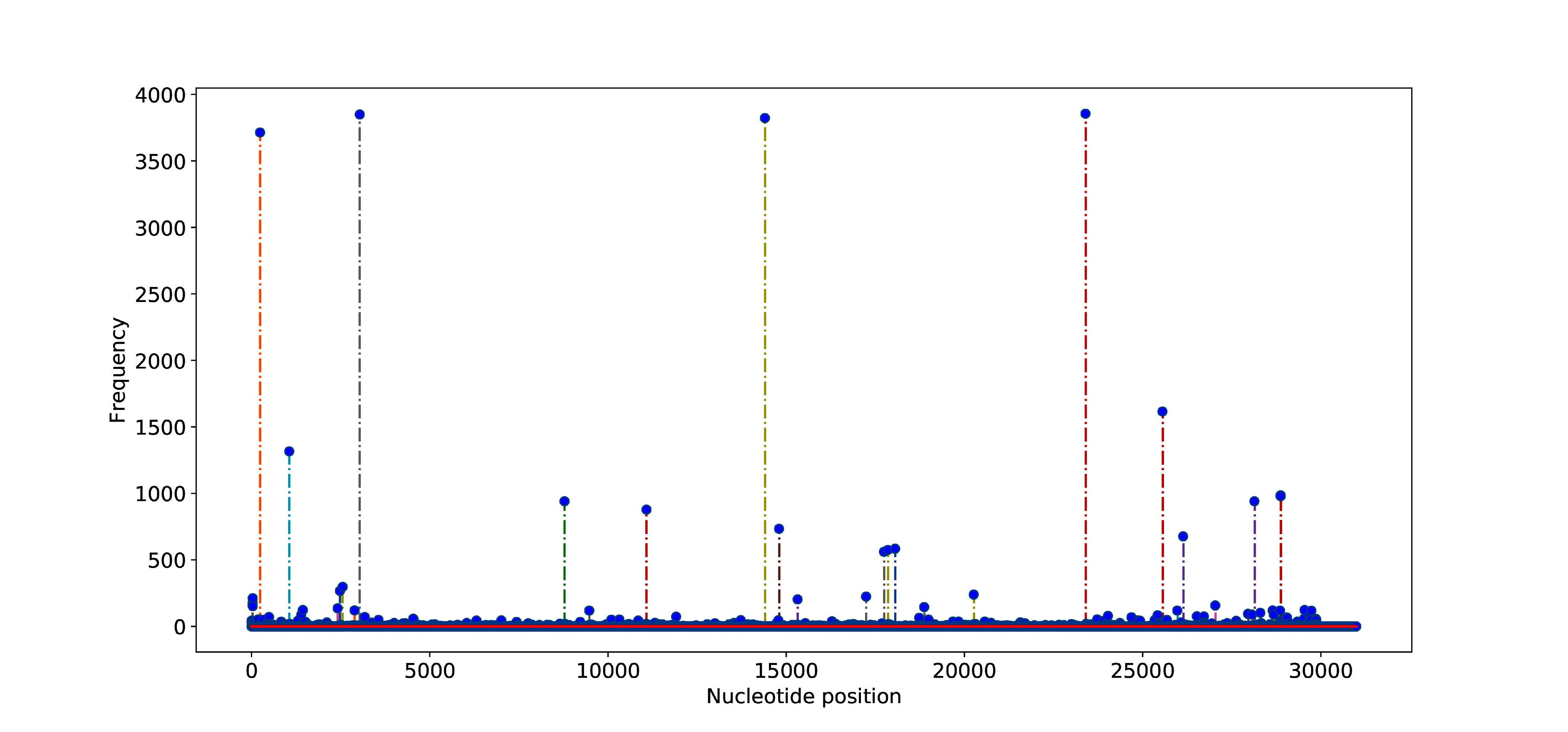}
\caption{Frequencies of the single SNP mutations of SARS-CoV-2 on the genomes depicted in Fig. \ref{fig:subfig:Distribution}.}
\label{fig:Distribution_Frequencies}
\end{figure}

\subsection{Ramification on COVID-19 diagnosis, vaccine and medicine }

\subsubsection{Protein-specific mutation analysis}

\begin{table}[H]
    \centering
    \setlength\tabcolsep{5pt}
    \captionsetup{margin=0.9cm}
    \caption{Protein-specific statistics of SARS-CoV-2 single mutations. Length refers to the number of codons in the genome associated with a specific protein.}
    \begin{tabular}{lccccc}
    \hline
    Protein &  Length & \# of  mutations &  mutation ratio & Mutation $h$-index  \\
    \hline
	Spike protein               & 1273  & 385  &0.30   &16   \\
   Main protease               & 306    & 68    & 0.22  & 9    \\
    Papain-like protease    & 1945  & 599  & 0.31  & 15   \\
    RNA polymerase          & 932    & 223  & 0.24  & 13   \\
    Endoribo-nuclease       & 346    & 87    &0.25   & 9    \\
    Envelope protein         & 75      & 13    &0.17   &5    \\
    Membrane protein       & 222    & 63    &0.28   &9    \\
    Nucleocapsid protein  & 419    & 235  &0.56   &27   \\
    \hline
    \end{tabular}
    \label{table:hindex}
\end{table}

Table \ref{table:hindex} presents the statistics of single mutations on various SARS-CoV-2 proteins that occurred in the recorded genomes between January 5, 2020, and April 24, 2020.  The papain-like protease has the highest number of mutations of 599 while the envelope protein has the lowest number of mutations of 13.  Since the sizes of proteins vary dramatically from 1945 for the papain-like protease to 75 for the envelope protein, it is useful to consider the mutation ratio, i.e., the number of mutations per residue. In this category, the envelope protein still has the lowest score of 0.17, whereas the nucleocapsid protein has the highest score of 0.56, i.e, 235 mutations on its 419 residues.  Note that 3CL protease has the second-lowest   mutation ratio of 0.22, indicating its conservative nature. Another relatively conservative protein judged by the  mutations ratio is the RNA-dependent RNA polymerase. It has 223 mutations over its 932 residues. 

Counting the number of single mutations and mutation ratio does not reflect the fact some mutations occur numerous times over genome samples while other mutations may happen only on a few genome samples. To account for the frequency effect of mutations, we introduce a mutation $h$-index to measure both the number of mutations and the frequency of mutations of a given protein or genetic section. It is defined as the maximum value of $h$ such that the given protein  genetic section has $h$ single mutations that have each occurred at least $h$ times.  It is very interesting to note from Table \ref{table:hindex} that the mutation $h$-index correlates very well with the number of mutations per residue. Specifically,  nucleocapsid protein has both the highest number of mutations per residues of 0.56 and the highest $h$-index of 27, suggesting that it is the most non-conservative protein in SARS-CoV-2 genomes. In contrast, the envelope protein has the lowest number of mutations per residues of 0.17 and the lowest $h$-index of 5, indicating its relatively conservative nature. By combining the number of mutations per residue and the mutation $h$-index, we report that the three most conservative  SARS-CoV-2 proteins are 1) the envelope protein, 2) the main protease, and 3) the endoribonuclease. It is found that the most non-conservative SARS-CoV-2 proteins are 1) the nucleocapsid protein, 2) the spike protein and 3) the papain-like protease. 

\subsubsection{Diagnosis}
Real-time RT-PCR (rRT-PCR) is routinely used in the qualitative detection of nucleic acid from SARS-CoV-2 for diagnostic testing COVID-19 \cite{corman2020detection,udugama2020diagnosing}. The primers used in the rRT-PCR are critical for the precise diagnosis of COVID-19 and the discovery of new strains. The primer sequences are specially designed for amplifying the conserved regions across the different existing strains for high specificity and sensitivity, and also are subject to genotype changes as the SARS-CoV-2 coronavirus evolves. In diagnostic testing COVID-19, many rRT-PCR primers are designed to detect for three perceived conservative SARS-CoV-2 regions: (1) RNA-dependent RNA polymerase (RdRP) gene in ORF1ab region, (2) the E protein gene, and (3) the N protein gene \cite{corman2020detection}. Our genotyping statistics given in Table \ref{table:hindex} indicates that the nucleocapsid protein is the worst choice. 

 Among four structural proteins of SARS-CoV-2, the spike surface glycoprotein (S) of 1273 amino acid residues, nucleocapsid protein (N) of 419 amino acid residues, membrane protein (M) of 222 amino acid residues, envelope protein (E) of 75 amino acid residues, the S protein is the most divergent with 385 unique mutations among the 6156 SARS-CoV-2 genomes. The N protein has 235 unique mutations, the E protein has 13 mutations. Considering the lengths of the proteins, all the four structural proteins undergo high mutations. The RdRP gene, which is often used in diagnostic testing COVID-19, also has 223 mutations.

Therefore, all the three regions in routine rRT-PCR target, namely RdRP, the N protein gene, and the E protein genes, have significant mutations. Precise and robust diagnosis tools must be re-established according to the conserved regions and predominated mutations in the SARS-CoV-2 genomes detailed in the Supporting Material.  

\subsubsection{Vaccine development}

Notably, SARS-CoV-2 has a unique furin cleavage site, where four amino acid residues (PRRA) are inserted into the S1-S2 junction region 681-684 of the S protein \cite{walls2020structure}. The furin cleavage site is crucial for zoonotic transmission of SARS-CoV-2 \cite{follis2006furin}. This study reveals crucial mutations near the S1-S2 junction region in the S protein, including 23403A$>$G-(D614G), 23422C$>$T-(V620V), 23575C$>$T-(C671C), 23586A$>$G-(Q674R), 23611G$>$A-(R683R), 23707C$>$T-(P715P), 23731C$>$T-(T723T), 23849T$>$C-(L763L), and 23929C$>$T-(Y788Y). Moreover, these mutations of the S protein SARS-CoV-2 are located at the epitope region, corresponding to the regions 469-882 and 599-620 in SARS-CoV) \cite{ren2003strategy}.

Additionally, many mutated amino acids are on the surface of the S protein as shown in  Fig. \ref{fig:Spike}. 
Unfortunately, the S protein is the second most non-conservative protein in the genome based on the number of mutations per residue and mutation $h$-index. In fact, about half of the receptor-binding domain residues of the S proteins have had mutations in the past few months as shown in Fig. \ref{fig:Spike2}. Because the surface accessibility of epitope is also important for the interaction of antibody and antigen, these mutations are critical for the antigenicity of the S protein. 

The convalescent COVID-19 patients show a neutralizing antibody response after infection, which are directed against the S protein or the N protein \cite{raoult2020coronavirus}. The neutralizing antibody responses against SARS-CoV-2 could give some defense against SARS-CoV-2 infection and thus, having implications for preventing SARS-CoV-2 outbreaks. The divergence of spike proteins, the non-conserved regions of the spike proteins might contribute to the antigenicity. The high frequent mutations identified in the S protein and the N proteins must be considered when designing a vaccine. 
 
\subsubsection{Drug discovery}
 Unfortunately, there is no specific effective drug for  SARS-CoV-2 at this point. Much of the drug discovery effort focuses on  SARS-CoV-2 non-structural proteins.   Among the major non-structural proteins of SARS-CoV-2, the main protease of 306 amino acids has  68 mutations with 0.22 mutations per residue and the mutation $h$-index of 9, RNA polymerase of 932 amino acids has 223 mutations with 0.24 mutations per residue and the mutation $h$-index of 13, and papain-like protease of 1945 amino acids has 599 mutations with 0.31 mutations per residue and the mutation $h$-index of 15. In fact, the main protease is the most popular drug target because there are no similar known genes in the human genome, which implies  SARS-CoV-2 main protease inhibitors will be likely less  toxic \cite{jin2020structure}. The present study suggests that the main protease is the second most conservative protein. Therefore, it remains the most attractive target for drug discovery.

\subsubsection{Protein-specific discussion}

\paragraph{Spike glycoprotein }

\begin{table}[H]
    \centering
    \setlength\tabcolsep{2pt}
    \captionsetup{margin=0.9cm}
    \caption{Top 10 high frequency single SNP genotypes in the spike surface glycoprotein of SARS-CoV-2.}
    \begin{tabular}{lcccccccc}
    \hline
     Rank     &SNP position & Protein mutation  & Total frequency & Cluster I  & Cluster II & Cluster III  & Cluster IV  & Cluster V  \\ \hline
     Top1        & 23403  & D614G   & 3897  & 1  & 86    & 1802 & 963  & 1045  \\
     Top2        & 23731  & T723T   & 53    & 0  & 1     & 0   & 52   & 0     \\
     Top3        & 24862  & T1100T  & 45    & 0  & 0     & 45  & 0    & 0     \\
     Top4        & 23929  & Y789Y   & 43    & 0  & 43    & 0   & 0    & 0     \\
     Top5        & 21575  & L4F     & 30    & 4  & 11    & 6   & 4    & 5     \\ 
     Top6        & 24368  & D936Y   & 26    & 0  & 2     & 6   & 0    & 18    \\
     Top7        & 21707  & H49Y    & 24    & 4  & 20    & 0   & 0    & 0     \\
     Top8        & 23707  & P715P   & 23    & 0  & 18    & 1   & 0    & 0     \\
     Top9        & 23422  & V620V   & 20    & 4 & 4    & 2  & 14    & 0     \\
     Top10       & 23010  & V483A   & 18    & 18 & 0     & 0   & 0    & 0     \\\hline
    \end{tabular}
    \label{table:spike}
\end{table}

\begin{figure}[H]
\centering
\includegraphics[scale=0.20]{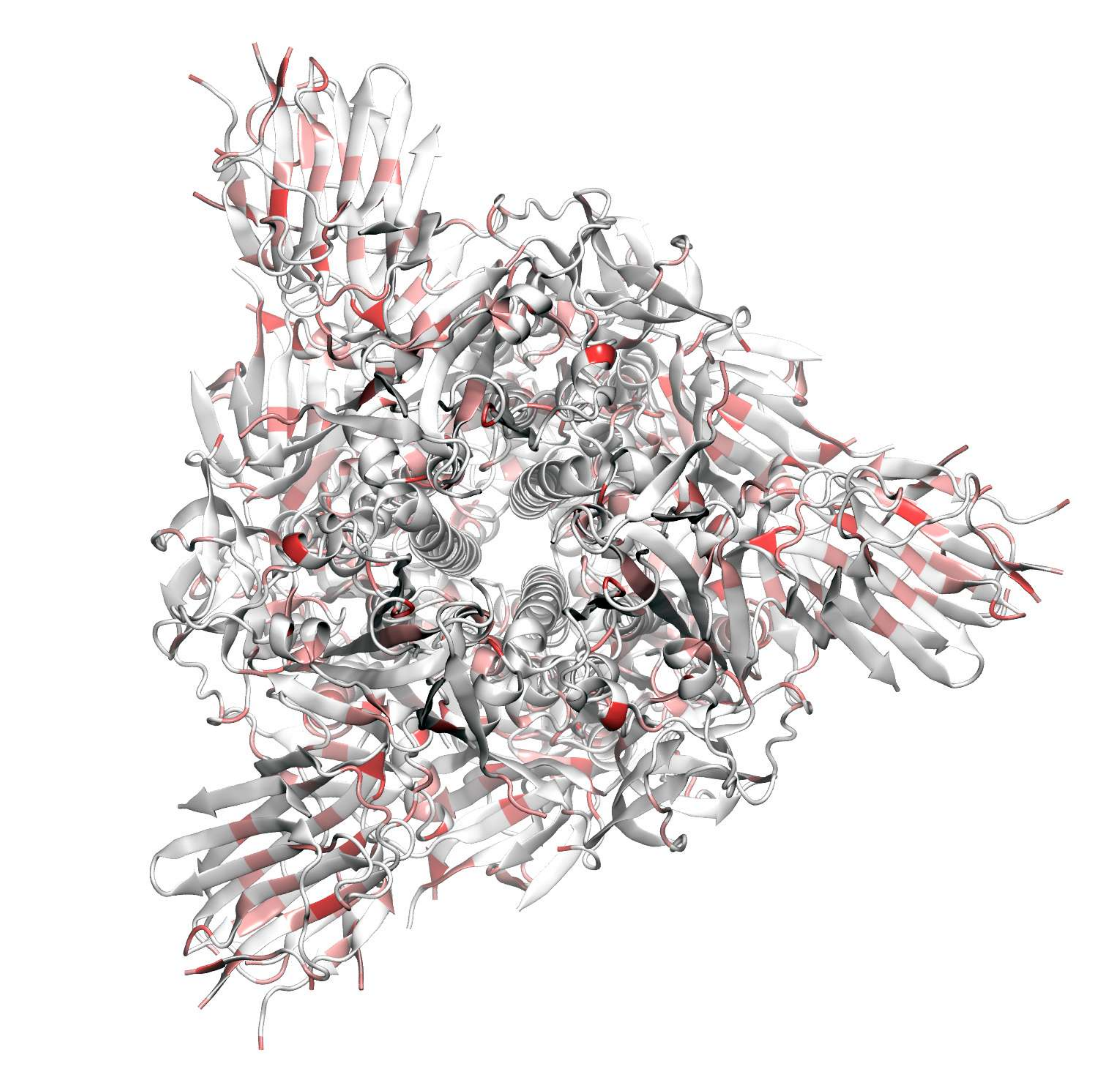}
\caption{Illustration of SARS-CoV-2 spike protein mutations using 6VXX as a template  \cite{walls2020structure}.}
\label{fig:Spike}
\end{figure}

\begin{figure}[ht]
\centering
\includegraphics[scale=0.15]{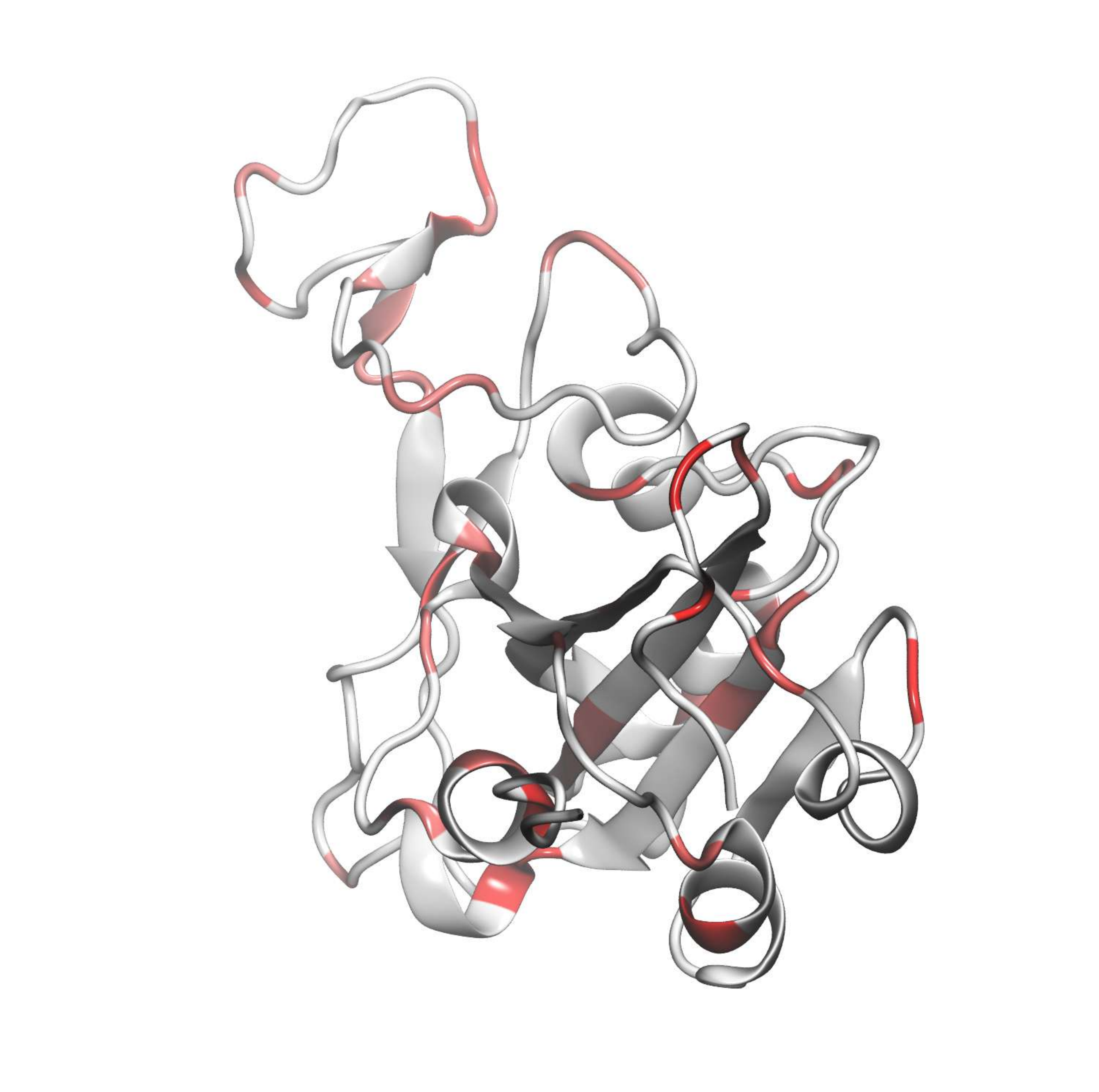}
\caption{Illustration of SARS-CoV-2 spike-protein receptor binding domain (RBD) mutation using 6m0j as a template. It is noted that near half of residues in the RBD have undergone mutations in the few months.}
\label{fig:Spike2}
\end{figure}

The SARS-CoV-2 spike glycoprotein, or S protein, comprised of two subunits, S1 and S2, of very different properties \cite{walls2020structure}, see Fig. \ref{fig:Spike}. Among them,  the S1 subunit, as shown in Fig. \ref{fig:Spike},  contains the receptor-binding domain (RBD) responsible for binding to the host cell receptor angiotensin-converting enzyme 2 (ACE2).  The RBD is also the common binding domain for antibodies. The S2 subunit offers the structural support of the S protein and mediates fusion between the viral and host cell membranes. After the fusion, the virus releases the viral genome into the host cell. 
 
 The S1 RBD protein plays key parts in the induction of neutralizing-antibody and T-cell responses, as well as protective immunity. However, S2 and extracellular domain (ECD) of spike protein and their combination are commonly used in recombinant proteins in SARS-CoV-2 antibody development.  
 
As shown in Table \ref{table:hindex}, the S protein is the most heterogeneous structural protein with a significant number of mutations as shown in Figs. \ref{fig:Spike} and \ref{fig:Spike2} and  Table \ref{table:spike}. The divergence of the spike protein, the non-conserved regions of the spike protein might contribute to the antigenicity difference in SARS-CoV-2 isolates. We found that most of the high frequent mutations of the S protein are located in the S1 subunit. Figure \ref{fig:Spike2} indicates that near half of the amino acid residues have had mutations since January 5, 2020.  One of the important mutations at S1 is 23010 (V483A) within the RBD for ACE2 binding. The structural study revealed that the amino acids 442-487 in the S1 subunit may impact viral binding to human ACE2 \cite{hoffmann2020sars,wan2020receptor}. The mutations identified in this study imply the change in ACE2 binding affinity and the transmissibility of SARS-CoV-2  as well as negative impacts in preventive vaccine and diagnostic test development.

\paragraph{Main protease}

\begin{table}[H]
    \centering
    \setlength\tabcolsep{2pt}
    \captionsetup{margin=0.9cm}
\caption{Top 11 high frequency single SNP genotypes in the main protease of SARS-CoV-2.}
    \begin{tabular}{lcccccccc}
    \hline
     Rank     &SNP position & Protein mutation  & Total frequency & Cluster I  & Cluster II & Cluster III  & Cluster IV  & Cluster V  \\ \hline
     Top1        & 10323  & K90R   & 52    & 1  & 8     & 43   & 0    & 0     \\
     Top2        & 10097  & G15S   & 51    & 0  & 1     & 0    & 50   & 0     \\
     Top3        & 10851  & A266V  & 44    & 0  & 2     & 16   & 0    & 26    \\
     Top4        & 10582  & D176D  & 19    & 0  & 0     & 5    & 0    & 14    \\
     Top5        & 10771  & Y239Y  & 15    & 15 & 0     & 0    & 0    & 0     \\ 
     Top6        & 10507  & N151N  & 11    & 0  & 10    & 1    & 0    & 0     \\
     Top7        & 10948  & R298R  & 11    & 0  & 0     & 0    & 11   & 0     \\
     Top8        & 10265  & G71S   & 9     & 0  & 0     & 0    & 9    & 0     \\
     Top9        & 10870  & L272L  & 9     & 0  & 0     & 1    & 4    & 4     \\
     Top10       & 10319  & L89F   & 8     & 0  & 1     & 4    & 0    & 3     \\
     Top11       & 10450  & P132P  & 8     & 0  & 0     & 8    & 0    & 0     \\
     \hline
    \end{tabular}
    \label{table:3C-like_proteinase}
\end{table}

\begin{figure}[ht]
\centering
\includegraphics[scale=0.15]{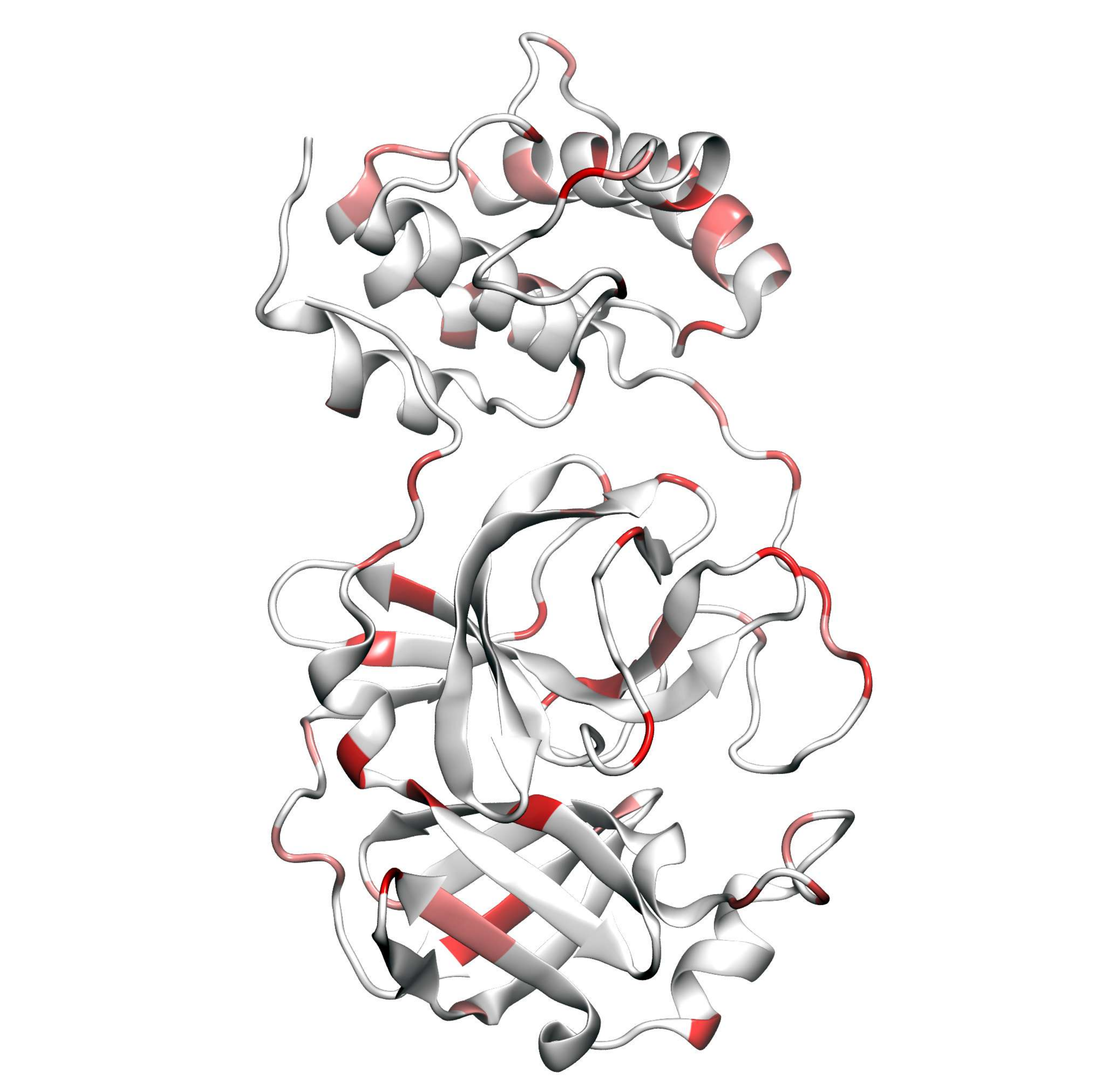}
\caption{Illustration of SARS-CoV-2 main protease mutations using 6LU7 as a template \cite{jin2020structure}.}
\label{fig:Mainprotein}
\end{figure}

\begin{figure}[ht]
\centering
\includegraphics[scale=0.15]{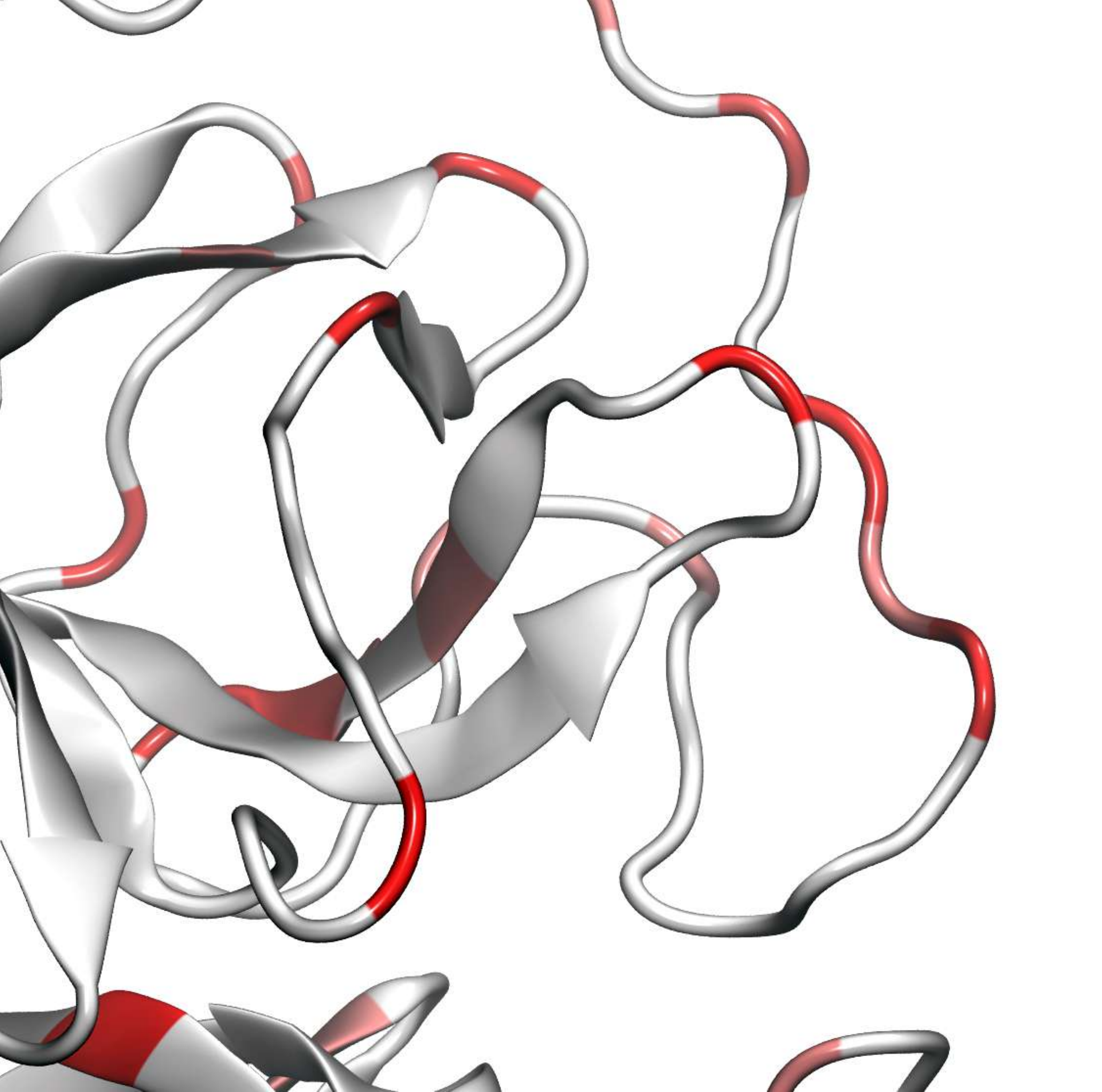}
\caption{Illustration of SARS-CoV-2 main protease binding domain (BD) mutations of 6LU7.}
\label{fig:MainproteinBD}
\end{figure}
SARS-CoV-2 main protease, or 3CL protease, is essential for cleaving the polyproteins that are translated from the viral RNA \cite{jin2020structure}. It operates at multiple cleavage sites on the large polyprotein through the proteolytic processing of replicase polyproteins and plays a pivotal role in viral gene expression and replication. SARS-CoV-2 main protease is one of the most attractive targets for anti-CoV drug design because its inhibition would block viral replication and it is unlikely to be toxic due to no known similar human proteases.  Another reason for the focused drug discovery efforts in developing SARS-CoV-2 main protease inhibitors is that this protein is relatively conservative as shown in Table \ref{table:hindex}.
 
Figure \ref{fig:Mainprotein} illustrates the main protease mutation patterns. Figure \ref{fig:MainproteinBD} further highlights the inhibitor binding domain (BD). Indeed, the main protease is relatively conservative compared to the spike protein.  Table \ref{table:3C-like_proteinase} lists top 11 mutations and their frequency in our dataset. It is interesting to see that many mutations, such as Y239Y, N151N, R298R, L272L, and P132P, are degenerate ones. One possible explanation is that non-degenerate may be non-silent and likely cause unsurvivable disruption to the virus. Note that mutation G15S mostly occurs in Cluster IV. Mutation Y239Y is restricted to Cluster I.  Some other mutations, such as R298R, G71S, and P132P, are specific to certain clusters. Nonetheless, some mutations at the BD shown in Fig. \ref{fig:MainproteinBD} are worth noting. They can undermine the ongoing drug discovery effort. 

 \paragraph{Papain-like protease}

\begin{table}[H]
    \centering
    \setlength\tabcolsep{2pt}
    \captionsetup{margin=0.9cm}
    \caption{Top 10 high frequency single SNP genotypes in the papain-like protease of SARS-CoV-2.}
    \begin{tabular}{lcccccccc}
    \hline
     Rank     &SNP position & Protein mutation  & Total frequency & Cluster I  & Cluster II & Cluster III  & Cluster IV  & Cluster V  \\ \hline
     Top1        & 3037  & F106F   & 3889  & 0  & 80    & 1800  & 964  & 1045  \\
     Top2        & 2891  & A58T    & 120   & 0  & 119   & 0     & 1    & 0     \\
     Top3        & 3177  & P153L   & 72    & 0  & 69    & 2     & 1    & 0     \\
     Top4        & 4540  & Y607Y   & 60    & 0  & 60    & 0     & 0    & 0     \\
     Top5        & 7011  & A1431V  & 45    & 0  & 43    & 2     & 0    & 0     \\ 
     Top6        & 6312  & T1198K  & 44    & 0  & 42    & 1     & 1    & 0     \\
     Top7        & 7438  & Y1573Y  & 34    & 0  & 9     & 21    & 4    & 0     \\
     Top8        & 3373  & D218D   & 29    & 0  & 3     & 0     & 26   & 0     \\
     Top9        & 4002  & T428I   & 26    & 0  & 1     & 0     & 25   & 0     \\
     Top10       & 6040  & F1107F  & 26    & 0  & 10    & 12    & 0    & 4     \\\hline
    \end{tabular}
    \label{table:nsp3}
\end{table}

\begin{figure}[ht]
\centering
\includegraphics[scale=0.15]{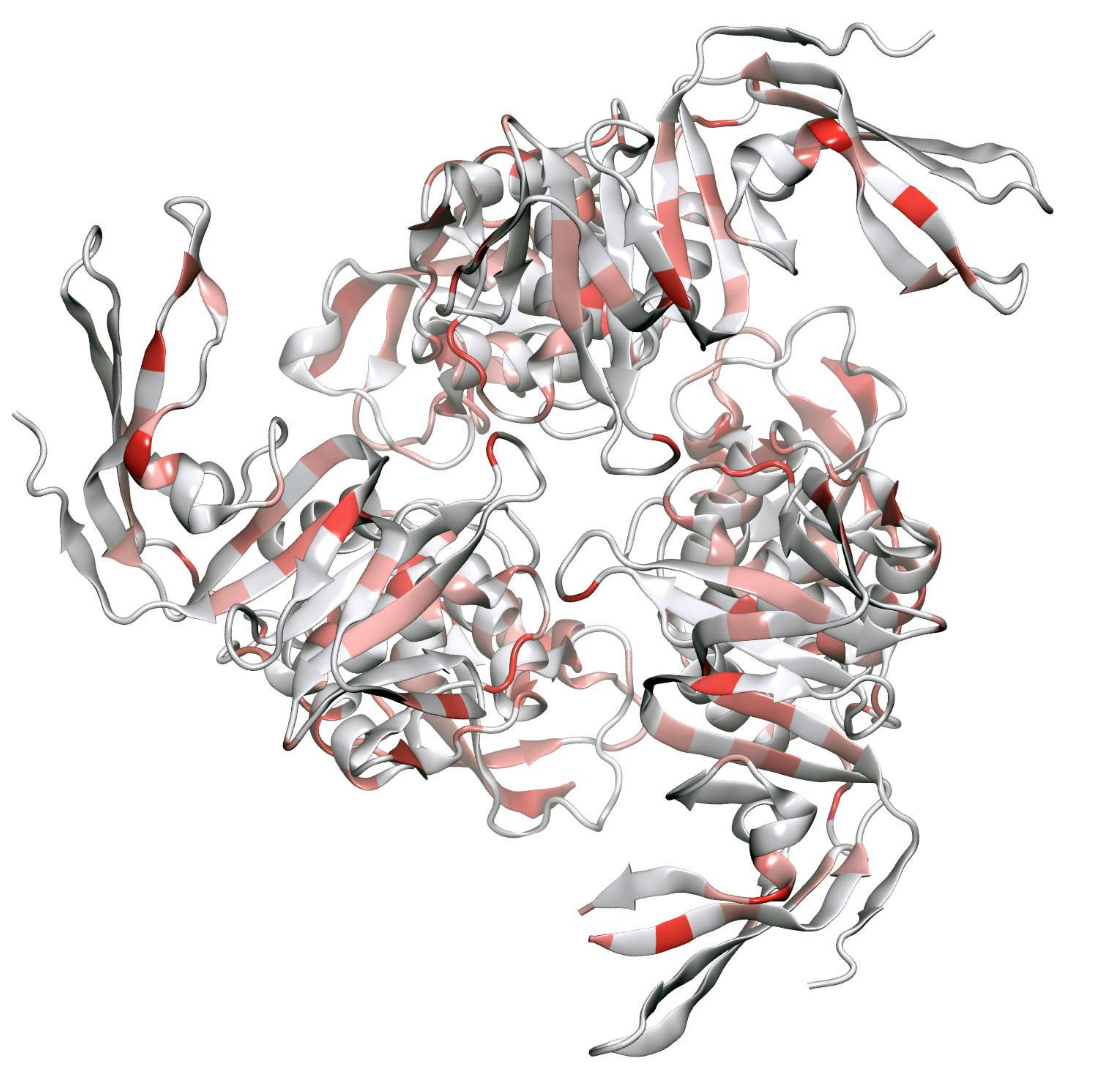}
\caption{Illustration of SARS-CoV-2 papain-like protease mutations using 6W9C as a template \cite{osipiuk2020crystal}.}
\label{fig:PL-protein-6W9C}
\end{figure}

SARS-CoV-2  papain-like protease (PLPro) is a cysteine cleavage protein located within the non-structural protein 3 (NS3) section of the viral genome \cite{osipiuk2020crystal}.  Like, the main protease, PLPro activity is required to cleave the viral polyprotein into functional, mature subunits and, thereby, contributes to the biogenesis of the virus replication. Additionally, PLpro possesses a deubiquitinating activity. The SARS PLPro is also a major therapeutic and diagnostic target. 

As shown in Table \ref{table:hindex},  the SARS PLPro is prone to mutations. Figure \ref{fig:PL-protein-6W9C} shows that mutations are all over the places in PLPro. Table \ref{table:nsp3} lists top ten mutations in PLPro. Five of these mutations are degenerate ones, including one of the highest frequented mutations.  Note that none one of the top mutations occurred in Cluster I. On the contrast, Cluster II has many different mutations. 

\paragraph{RNA polymerase}
 
\begin{table}[H]
    \centering
    \setlength\tabcolsep{2pt}
    \captionsetup{margin=0.9cm}
    \caption{Top 10 high frequency single SNP genotypes in the RNA dependent polymerase of SARS-CoV-2.}
    \begin{tabular}{lcccccccc}
    \hline
     Rank     &SNP position & Protein mutation  & Total frequency & Cluster I  & Cluster II & Cluster III  & Cluster IV  & Cluster V  \\ \hline
     Top1        & 14408  & P323L   & 3859  & 0  & 76    & 1785   & 954  & 1044  \\
     Top2        & 14085  & Y455Y   & 739   & 1  & 728   & 6      & 4    & 0     \\
     Top3        & 15324  & N628N   & 204   & 0  & 10    & 189    & 4    & 1     \\
     Top4        & 13730  & A96V    & 47    & 0  & 43    & 4      & 0    & 0     \\
     Top5        & 14786  & A449V   & 46    & 0  & 3     & 30     & 10   & 3     \\ 
     Top6        & 13536  & Y32Y    & 26    & 0  & 1     & 0      & 25   & 0     \\
     Top7        & 13627  & D63Y    & 26    & 0  & 25    & 1      & 0    & 0     \\
     Top8        & 15540  & V700V   & 25    & 0  & 25    & 0      & 0    & 0     \\
     Top9        & 13568  & A42V    & 17    & 0  & 0     & 17     & 0    & 0     \\
     Top10       & 14073  & D211D   & 15    & 0  & 0    & 15     & 0    & 0     \\\hline
    \end{tabular}
    \label{table:RNAdependentpolymerase}
\end{table}

\begin{figure}[ht]
\centering
\includegraphics[scale=0.15]{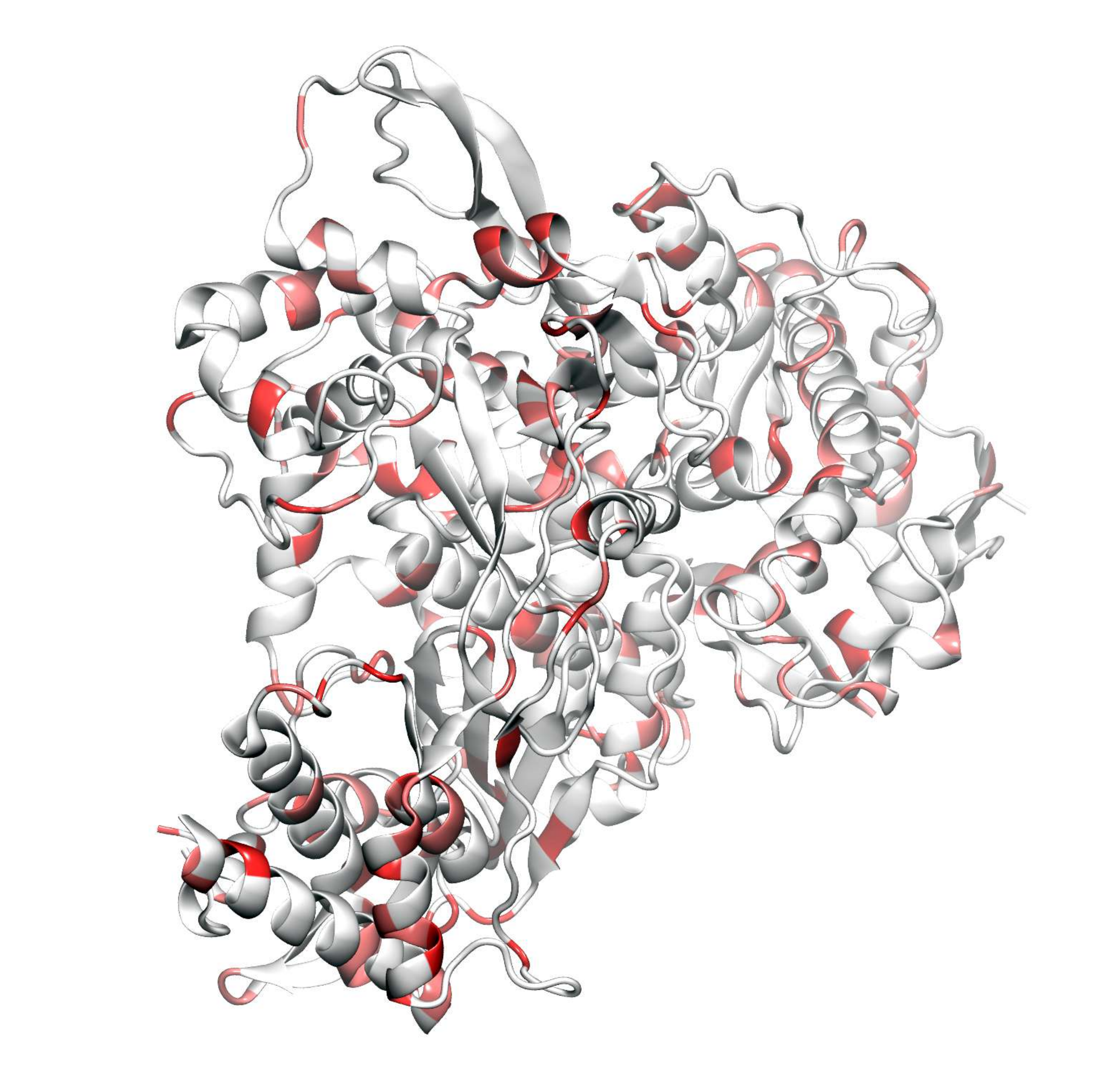}
\caption{Illustration of SARS-CoV-2 RNA-polymerase mutations  using  6M71 as a template  \cite{gao2020structure}.}
\label{fig:RNA-polymerase-6m71}
\end{figure}

 SARS RNA-dependent RNA polymerase (RdRP) is an enzyme that catalyzes the synthesis of the SARS RNA strand complementary to the SARS-CoV-2 RNA template and thus essential to the replication of SARS-CoV-2  RNA  \cite{gao2020structure}. As one of the non-structure proteins, RdRPs located in the early part of ORF1b section.  Like most other RNA viruses, SARS-CoV-2 RdRPs are considered to be highly conserved to maintain viral functions and thus targeted in antiviral drug development as well as diagnostic tests. On the other hand,  the SARS-CoV-2 RNA polymerase lacks proofreading capability and thus its mutations are deemed to happen as shown in Table \ref{table:hindex}.

Figure \ref{fig:RNA-polymerase-6m71} illustrates the SARS-CoV-2 RdRP mutations since January 5, 2020. Surprisingly, there are many mutations in SARS-CoV-2 RdRP. Table \ref{table:RNAdependentpolymerase} describes the top ten mutations.  As in other cases, five of these mutations are degenerate ones. Cluster I has no nondegenerate mutations. 

\paragraph{Endoribo-nuclease}

\begin{table}[H]
    \centering
    \setlength\tabcolsep{2pt}
    \captionsetup{margin=0.9cm}
    \caption{Top 12 high frequency single SNP genotypes in the endoribo-nuclease of SARS-CoV-2.}
    \begin{tabular}{lcccccccc}
    \hline
     Rank     &SNP position & Protein mutation  & Total frequency & Cluster I  & Cluster II & Cluster III  & Cluster IV  & Cluster V  \\ \hline
     Top1        & 19684   & V22L   & 38    & 0  & 37    & 0   & 1    & 0    \\
     Top2        & 19839   & N73N   & 37    & 0  & 0     & 0   & 37   & 0    \\
     Top3        & 20578   & V320L  & 37    & 0  & 1     & 36  & 0    & 0    \\
     Top4        & 20316   & F232F  & 19    & 0  & 19    & 0   & 0    & 0    \\
     Top5        & 20275   & D219Y  & 17    & 0  & 2     & 13  & 1    & 1    \\ 
     Top6        & 20031   & A137A  & 12    & 0  & 0     & 0   & 12   & 0    \\
     Top7        & 20005   & G129S  & 10    & 0  & 0     & 4   & 0    & 6    \\
     Top8        & 20134   & V172L  & 10    & 0  & 8     & 0   & 0    & 2    \\
     Top9        & 20270   & A217V  & 9     & 0  & 9     & 0   & 0    & 0    \\
     Top10       & 19961   & T114K  & 6     & 0  & 5     & 0   & 1    & 0    \\
     Top11       & 19862   & A81V   & 6     & 0  & 5     & 0   & 0    & 1    \\
     Top12       & 19645   & V9F    & 6     & 6  & 0     & 0   & 0    & 0    \\
     \hline
    \end{tabular}
    \label{table:endoribonuclease}
\end{table}

\begin{figure}[ht]
\centering
\includegraphics[scale=0.15]{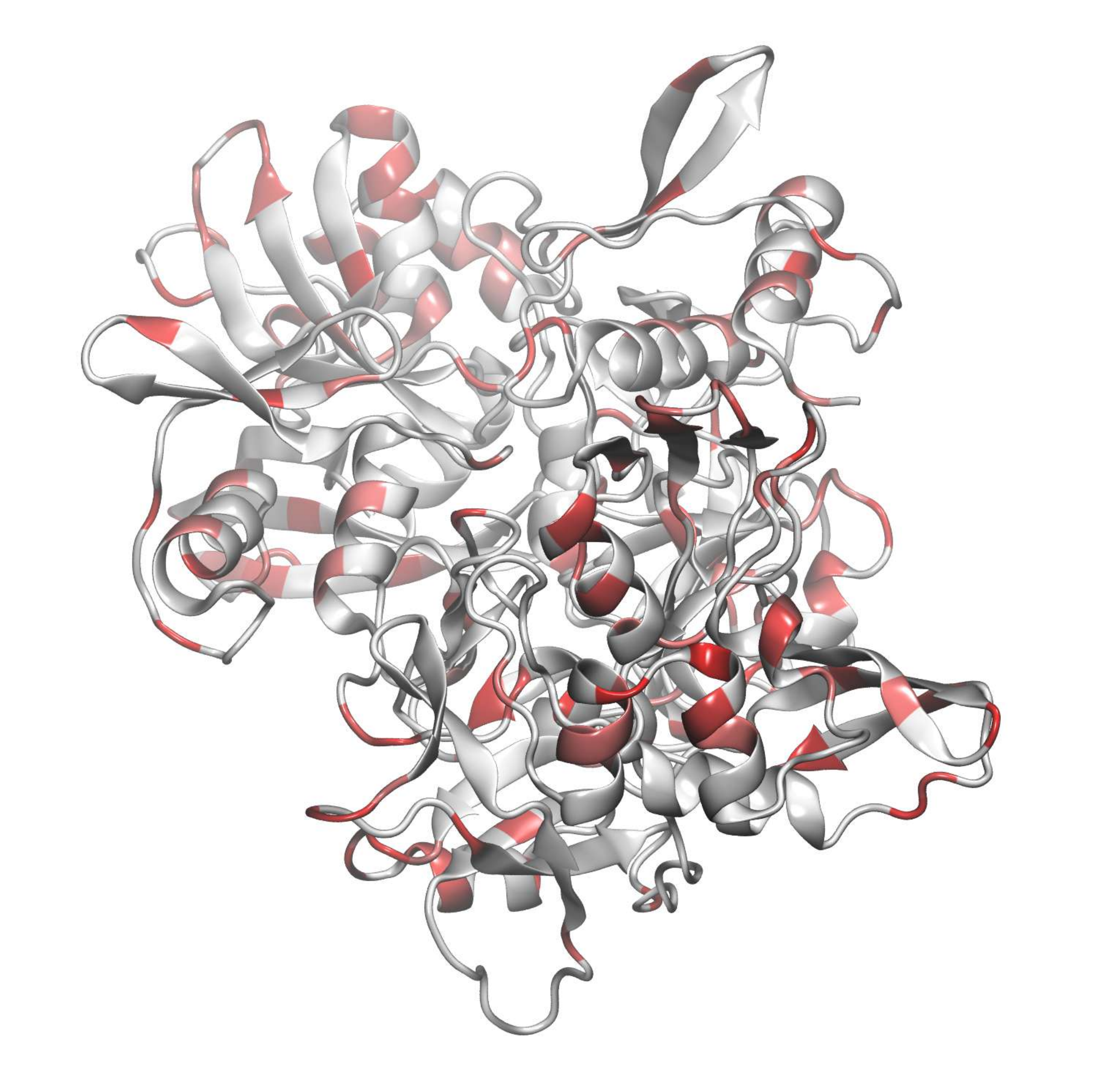}
\caption{Illustration of SARS-CoV-2  Endoribo-nuclease  protein mutations using 6VWW as a template  \cite{kim2020crystal}.}
\label{fig:Endo-protein-6VWW}
\end{figure}
Endoribo-nuclease (NendoU) protein is a nidoviral RNAuridylate-specific enzyme that cleaves RNA \cite{kim2020crystal}. It contains a C-terminal catalytic domain belonging to the EndoU family  RNA processing. The NendoU protein is presented among coronaviruses, arteriviruses, and toroviruses. The many aspects of the detailed function and activity of SARS-CoV-2 NendoU protein are yet to be revealed. 

Figure \ref{fig:Endo-protein-6VWW} depicts SARS-CoV-2 NendoU protein mutations.  Like in most other SARS-CoV-2 proteins, mutations have occurred over different parts. Table \ref{table:hindex} shows that NendoU is relatively conservative. Table \ref{table:endoribonuclease} lists the top twelve high-frequency mutations of the SARS-CoV-2 NendoU protein that occurred in the past few months. Three of these mutations are degenerate ones. The frequencies of these mutations range from 38 to 6. Note that Cluster I do not have any of these mutations. 

\paragraph{Envelope protein}

\begin{table}[H]
    \centering
    \setlength\tabcolsep{2pt}
    \captionsetup{margin=0.9cm}
    \caption{Top 13 high frequency single SNP genotypes in the envelope (E) protein of SARS-CoV-2.}
    \begin{tabular}{lcccccccc}
    \hline
     Rank     &SNP position & Protein mutation  & Total frequency & Cluster I  & Cluster II & Cluster III  & Cluster IV  & Cluster V  \\ \hline
     Top1        & 26319  & V25V    & 8     & 0  & 7  & 1  & 0  & 0     \\
     Top2        & 26340  & A32A    & 7     & 0  & 5  & 2  & 0  & 0     \\
     Top3        & 26326  & L28L    & 5     & 0  & 5  & 0  & 0  & 0     \\
     Top4        & 26256  & F4F     & 4     & 0  & 0  & 2  & 2  & 0     \\
     Top5        & 26301  & L19L    & 3     & 0  & 2  & 1  & 0  & 0     \\ 
     Top6        & 26433  & K63K    & 2     & 0  & 0  & 1  & 0  & 1     \\
     Top7        & 26370  & Y42Y    & 1     & 0  & 1  & 0  & 0  & 0     \\
     Top8        & 26392  & S50G    & 1     & 0  & 1  & 0  & 0  & 0     \\
     Top9        & 26313  & F23F    & 1     & 0  & 0  & 1  & 0  & 0     \\
     Top10       & 26398  & V52I    & 1     & 0  & 0  & 0  & 0  & 1     \\
     Top11       & 26428  & V62F    & 1     & 0  & 0  & 0  & 1  & 0     \\
     Top12       & 26353  & L37F    & 1     & 0  & 1  & 0  & 0  & 0     \\
     Top13       & 26408  & S55F    & 1     & 0  & 0  & 1  & 0  & 0     \\ 
     \hline
    \end{tabular}
    \label{table:envelopeprotein}
\end{table}

\begin{figure}[ht]
\centering
\includegraphics[scale=0.15]{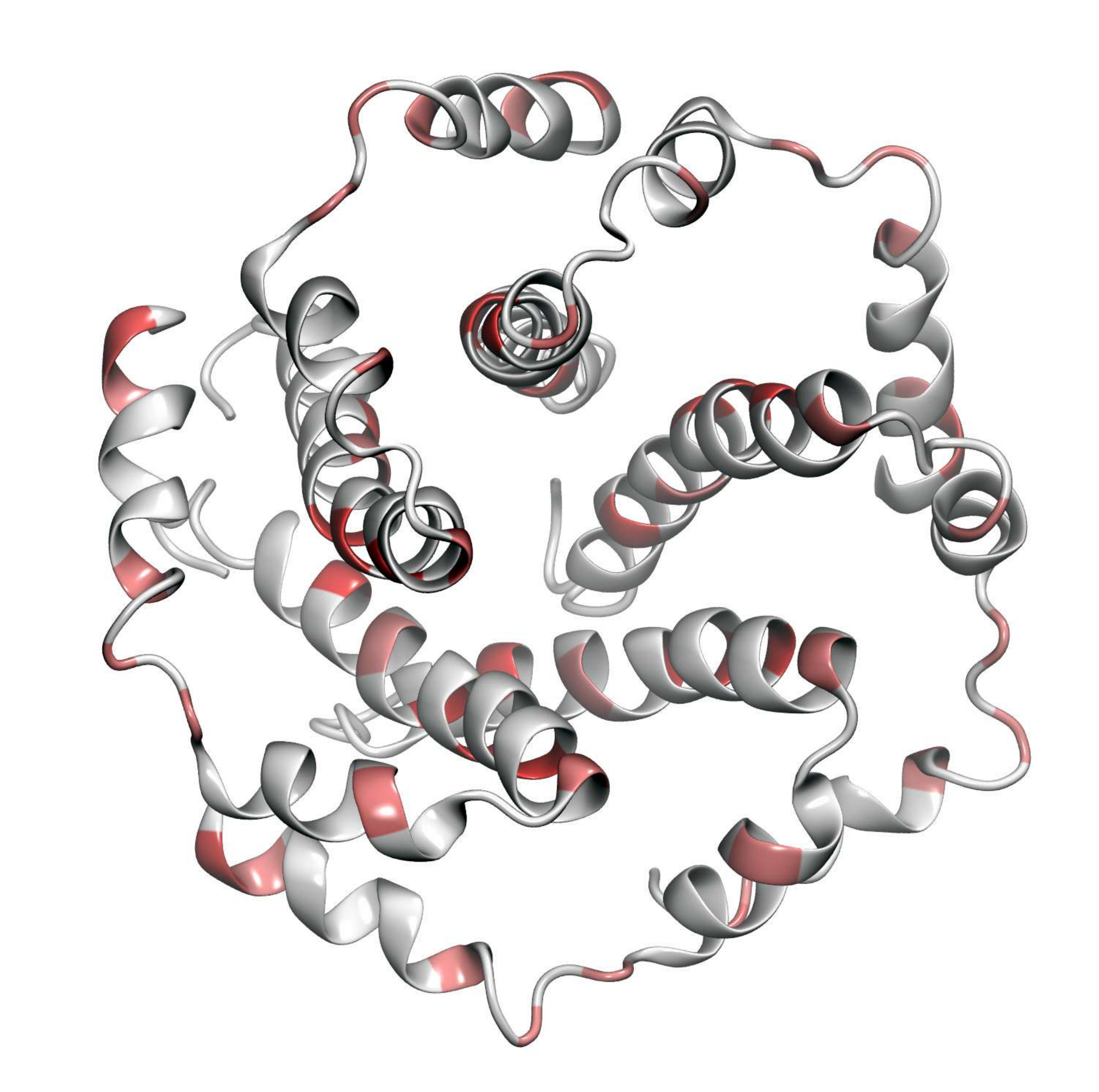}
\caption{Illustration of SARS-CoV-2  envelope protein mutations using  5X29 as template \cite{surya2018structural}.}
\label{fig:E-protein-5x29}
\end{figure}

The SARS-CoV-2 envelope (E) protein is one of SARS-CoV's four structural proteins. As a  transmembrane protein, it involves in ion channel activity,  and thus facilitates viral assembly, budding,  envelope formation, pathogenesis, and release of the virus  \cite{surya2018structural}. The E protein may not be essential for viral replication but it is for pathogenesis. 

Figure \ref{fig:E-protein-5x29} illustrates E protein as a very small pentamer with a few mutations. Table  \ref{table:envelopeprotein} shows its top thirteen mutations. Note that the first 7 mutations are degenerate ones. All other mutations have very low frequencies. As shown in Table \ref{table:hindex},  the SARS-CoV-2 E protein is very conservative.

\paragraph{Nucleocapsid protein}
\begin{table}[H]
    \centering
    \setlength\tabcolsep{2pt}
    \captionsetup{margin=0.9cm}
    \caption{Top 14 high frequency single SNP genotypes in the nucleocapsid phosphoprotein of SARS-CoV-2.}
    \begin{tabular}{lcccccccc}
    \hline
     Rank     &SNP position & Protein mutation  & Total frequency & Cluster I  & Cluster II & Cluster III  & Cluster IV  & Cluster V  \\ \hline
     Top1        & 28881  & R203K   & 989   & 1  & 20    & 4   & 964  & 0     \\
     Top2        & 28882  & R203R   & 983   & 0  & 18    & 1   & 964  & 0     \\
     Top3        & 28883  & G203R   & 983   & 0  & 18    & 1   & 964  & 0     \\
     Top4        & 28657  & D128D   & 125   & 1  & 124   & 0   & 0    & 0     \\
     Top5        & 28311  & P13L    & 102   & 0  & 101   & 1   & 0    & 0     \\ 
     Top6        & 28688  & L139L   & 91    & 0  & 90    & 1   & 0    & 0     \\
     Top7        & 29045  & P258T   & 67    & 0  & 65    & 2   & 0    & 0     \\
     Top8        & 29046  & P258R   & 67    & 0  & 65    & 2   & 0    & 0     \\
     Top9        & 29047  & P258P   & 67    & 0  & 65    & 2   & 0    & 0     \\
     Top10       & 29049  & R259L   & 67    & 0  & 65    & 2   & 0    & 0     \\
     Top11       & 29050  & R259R   & 67    & 0  & 65    & 2   & 0    & 0     \\
     Top12       & 29051  & Q260E   & 67    & 0  & 65    & 2   & 0    & 0     \\
     Top13       & 29052  & Q259R   & 67    & 0  & 65    & 2   & 0    & 0     \\
     Top14       & 29053  & Q260H   & 67    & 0  & 65    & 2   & 0    & 0     \\
     \hline
    \end{tabular}
    \label{table:nucleocapsidphosphoprotein}
\end{table}

\begin{figure}[ht]
\centering
\includegraphics[scale=0.15]{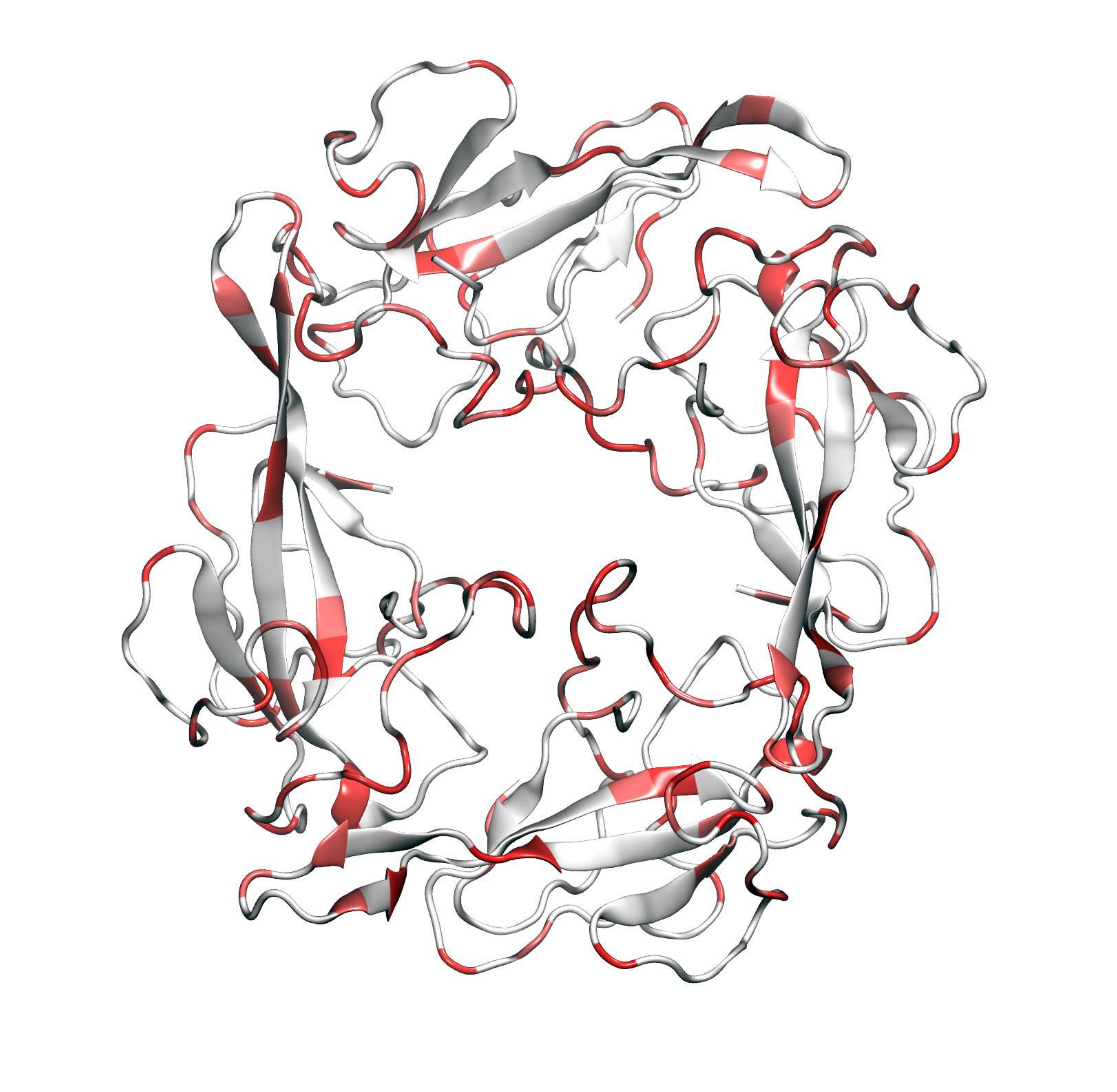}
\caption{Illustration of SARS-CoV-2 nucleocapsid phosphoprotein mutations using 6VYO as a template \cite{chang2020crystal}.}
\label{fig:N-protein-6VYO}
\end{figure}
SARS-CoV-2 nucleocapsid (N) protein is another structural protein. Its primary function is to encapsidate the viral genome. To do so, it is heavily phosphorylated (or charged) and thereby, can bind with RNA. Additionally, 
 SARS-CoV-2 N protein confirms the viral genome to replicase-transcriptase complex (RTC) and plays a crucial role in viral genome encapsulation. Therefore, it may function completely differently at different stages of the viral life cycle. SARS-CoV-2 N protein is considered to be one of the most conservative SARS-CoV-2 proteins in the literature and is a popular target for diagnosis of vaccine development \cite{corman2020detection}. The present works shown in Table \ref{table:hindex} indicates the SARS-CoV-2 N protein is the worst target of any drug, vaccine, and diagnostic development. 

Table \ref{table:nucleocapsidphosphoprotein} presents the top fourteen mutations of the SARS-CoV-2 N protein since January 5, 2020. Note that only three out of fourteen top mutations are degenerate ones, which is a significantly lower ratio than that of other proteins. The frequency of 14th mutation is 67, which suggests there are many mutations associated with these mediate-sized proteins. Most top mutations occurred to Clusters II and IV. Cluster V has none of the top fourteen mutations.   

\paragraph{Membrane protein}

\begin{table}[H]
    \centering
    \setlength\tabcolsep{2pt}
    \captionsetup{margin=0.9cm}
    \caption{Top 11 high frequency single SNP genotypes in the membrane glycoprotein of SARS-CoV-2.}
    \begin{tabular}{lcccccccc}
    \hline
     Rank    &SNP position & Protein mutation  & Total frequency & Cluster I  & Cluster II & Cluster III  & Cluster IV  & Cluster V  \\ \hline
     Top1        & 27046  & T175M   & 157  & 0  & 1   & 2   & 154  & 0     \\
     Top2        & 26729  & A69A    & 77   & 0  & 77  & 0   & 0    & 0     \\
     Top3        & 26530  & D3G     & 76   & 1  & 1   & 74   & 0    & 0     \\
     Top4        & 26951  & V143V   & 23   & 0  & 0   & 6   & 1    & 16    \\
     Top5        & 26750  & I76I    & 18   & 1  & 0   & 0   & 17   & 0     \\ 
     Top6        & 26864  & P114P   & 16   & 0  & 7   & 9   & 0    & 0     \\
     Top7        & 26730  & V70F    & 9    & 6  & 0   & 0   & 3    & 0     \\
     Top8        & 26625  & L35L    & 9    & 0  & 1   & 5   & 0    & 3     \\
     Top9        & 26681  & F53F    & 8    & 0  & 4   & 4   & 0    & 0     \\
     Top10       & 26720  & V66V    & 7    & 0  & 7   & 0   & 0    & 0     \\
     Top11       & 27005  & I161I   & 7    & 0  & 1   & 2   & 4    & 0     \\
     \hline
    \end{tabular}
    \label{table:membraneglycoprotein}
\end{table}

SARS-CoV-2 membrane (M) protein is another structural protein and plays a central role in viral assembly and viral particle formation. It exists as a dimer in the virion and has certain geometric shapes to enable certain membrane curvature and binding to nucleocapsid proteins. Similar to other SARS-CoV proteins, M protein is also a popular target for viral diagnosis and vaccines. 

Table \ref{table:hindex} gives SARS-CoV-2 M protein the meddle ranking for its conservation. Table \ref{table:membraneglycoprotein} details the top eleven mutations in SARS-CoV-2 M protein occurred in the past few months. Seven of these mutations are degenerate. Clusters I and V have relatively a few of these mutations. 

\section{Material and Methods}

\subsection{Data collection and pre-processing}
On January 5, 2020, the complete genome sequence of SARS-CoV-2 was first released on GenBank (Access number: NC\_045512.2) by Zhang's group at Fudan University \cite{wu2020new}. Since then, there has been a rapid accumulation of SARS-CoV-2 genome sequences. In this work, 6156 complete genome sequences with high coverage of SARS-CoV-2 strains from the infected individuals in the world are downloaded from the GISAID database \cite{shu2017gisaid} (\url{https://www.gisaid.org/}) as of April 24, 2020. All the records in GISAID without the exact submission date will not take into considerations. To rearrange the 6156 complete genome sequences according to the reference SARS-CoV-2 genome, multiple sequence alignment (MSA) is carried out by using Clustal Omega \cite{sievers2014clustal} with default parameters. 

\subsection{SNP genotyping}

SNP genotyping measures the genetic variations between different members of a species. Establishing the SNP genotyping method for the investigation of the genotype changes during the transmission and evolution of SARS-CoV-2 is of great importance. By analyzing the rearranged genome sequences, SNP profiles which record all of theSNP positions in teams of the nucleotide changes and its corresponding positions can be constructed. The SNP profiles of a given genome of a COVID-19 patient capture all the differences from a complete reference genome sequence and can be considered as the genotype of the individual SARS-CoV-2. 

\subsection{Distance of SNP variants}
The Jaccard distance measures dissimilarity between sample sets. The Jaccard distance of SNP variants is widely employed in the phylogenetic analysis of human or bacterial genomes \cite{yin2020genotyping}. In this work, we utilize the Jaccard distance to compare the difference between the SNP variant profiles of SARS-CoV-2 genomes. 

The Jaccard similarity coefficient,  also known as the Jaccard index, is defined as the intersection size divided by the union of two sets $A, B$ \cite{levandowsky1971distance}:
\begin{equation}
    J(A,B) = \frac{|A\cap B|}{|A\cup B|} = \frac{|A\cap B|}{|A|+|B|-|A\cap B|}.
\end{equation}
The Jaccard distance of two sets $A, B$ is scored as the difference between one and the Jaccard similarity coefficient and  is a metric on the collection of all finite sets:
\begin{equation}
    d_{J}(A,B) = 1 - J(A, B) = \frac{|A\cup B| - |A\cap B|}{|A\cup B|}.
\end{equation}
 Therefore, the genetic distance of two genomes corresponds to the Jaccard distance of their SNP variants.

In principle, the Jaccard distance measure of SNP variants takes account of the ordering ofSNP positions, i.e., transmission trajectory, when an appropriate reference sample is selected. However, one may fail to identify the infection pathways from the mutual Jaccard distances of multiple samples.  In this case,  the dates of the sample collections offer useful information. Additionally,  clustering techniques, such as $k$-means described below, enable us to characterize the spread of COVID-19 onto the communities.

\subsection{$K$-means clustering}
$K$-means clustering is one of the fundamental unsupervised algorithms in machine learning which aims at partitioning a given data set $X=\{x_1, x_2, \cdots, x_n, \cdots, x_N\}, x_n \in \mathbb{R}^d$ into $k$ clusters $\{C_1, C_2, \cdots, C_k\}, k \le N$ such that the specific clustering criteria are optimized. More specifically, the standard $K$-means clustering algorithm starts to pick $k$ points as cluster centers randomly and then allocates each data to its nearest cluster. The cluster centers will be updated iteratively by minimizing the within-cluster sum of squares (WCSS) which is defined by:
\begin{equation}
    \sum_{i=1}^{k}\sum_{x_i\in C_k}\|x_i-\mu_k\|_2^2,
\end{equation}
where $\mu_k=\_{i\in C_k}x_i/n_k$ is the mean of points located in the $k$-th cluster $C_k$ and $n_k$ is the number of points in $C_k$. Here, $\| \cdot \|_2$ denote the $L_2$ distance. 

 The algorithm above only provides a way to obtain the optimal partition for a fixed number of clusters. However, we are interested in finding the best number of clusters for the SNP variants. Therefore, the Elbow method is applied. By varying the number of clusters $k$, a set of WCSS can be calculated in the $K$-means clustering process, and then the plot of WCSS according to the number of clusters $k$ can be carried out. The location of the elbow in this plot will be considered as the optimal number of clusters. To be noticed, the WCSS measures the variability of the points within each cluster which is influenced by the number of points $N$. Therefore, as the number of total points of $N$ increases, the value of WCSS becomes larger. Additionally, the performance of $k$-means clustering depends on the selection of the specific distance.

In this work, we propose to implement $K$-means clustering with the Elbow method for analyzing the optimal number of the subtypes of SARS-CoV-2 SNP variants. The Jaccard distance-based and Location-based representations are considered as the input features for the $K$-means clustering method.

\subsubsection{Jaccard distance-based representation} Suppose we have a total of $N$ SNP variants concerning a reference genome in a SARS-CoV-2 sample. The location of the mutation sites for each SNP variant will be saved in the set $S_i, i=1,2,\cdots, N$. The Jaccard distance between two different sets (or samples) $S_i$, $S_j$ is denoted as $d_J(S_i, S_j)$. Therefore, the $N\times N$ Jaccard distance-based representation will be: 
\begin{equation}
    D_J(i,j) = d_J(S_i, S_j). 
\end{equation}

\subsubsection{Location-based representation} Suppose we have $N$ SNP variants with respect to a reference genome in a SARS-CoV-2 sample. Among them, $M$ different mutation sites can be counted. For $i$-th SNP variant, $V_i=[v_i^1, v_i^2, \cdots, v_i^M], i=1,2,\cdots, N$ is a $1\times M$ vector which satisfies:
\begin{equation}
v_i^j = 
\begin{cases}
1, \text{if  mutation happens at  location } $j$ \\
0, \text{otherwise.}
\end{cases}
\end{equation}
Therefore, an $N\times M$ location-based representation will be:
\begin{equation}
    L(i,j) = v_i^j
\end{equation}

\subsubsection{Principal component analysis (PCA)} Hundreds of complete genome sequences are deposit to the GISAID every day, which results in an ever-growing massive quantity of high dimensional data representations for the $K$-means clustering. For example, if the dataset of an organism involves 10,000 SNPs, the initial representation will be a 10,000-dimensional vector for each sample, which can be computationally difficult for a simple $K$-means clustering algorithm. Therefore, a dimensionality reduction method is used to pre-process the data.  The essential idea of PCA-based $K$-means clustering is to invoke the PCA to obtain a reduced-dimensional representation of each sample before performing the $K$-means clustering.  In practice, one can select a few lowest dimensional principal components as the $K$-means input for each sample.  In Ref. \cite{ding2004k}, the authors proved that the principal components are the continuous solution of the cluster indicators in the $K$-means clustering method, which provides us a rigorous mathematical tool to embed our high-dimensional data into a low-dimensional PCA subspace. 

\section{Conclusion}
The rapid global transmission of coronavirus disease 2019 (COVID-19) has offered some of the most heterogeneous, diverse, and challenging mutagenic environments to stimulate dramatic genetic evolution and response from severe acute respiratory syndrome coronavirus 2 (SARS-CoV-2). This work provides the most comprehensive genotyping of SARS-CoV-2 transmission and evolution up to date based on 6156 genome samples and reveals five clusters of the COVID-19 genomes and associated mutations on eight different SARS-CoV-2 proteins. We introduce mutation $h$-index and mutation ratio to qualify individual protein's degree of non-conservativeness. We unveil that SARS-CoV-2 envelope protein, main protease, and endoribonuclease protein are relatively the most conservative, whereas,  SARS-CoV-2 nucleocapsid protein,  spike protein, and papain-like protease are  relatively the most non-conservative. We report an alarming fact that all of the SARS-CoV-2 proteins have undergone intensive mutations since January 5, 2020, and some of these mutations may seriously undermine ongoing efforts on COVID-19 diagnostic testing, vaccine development, and drug discovery.

\section{Data Availability} The nucleotide sequences of the SARS-CoV-2 genomes used in this analysis are available, upon free registration, from the GISAID database (\url{https://www.gisaid.org/}). 
Eighteen tables are provided in the Supporting Material for SNP variants of 6156 SARS-CoV-2 samples across the world, SNP variants of 1625 SARS-CoV-2 samples in the US, SNP variants in five global clusters, SNP variants in three US clusters, and mutation records for eight SARS-CoV-2 proteins.  The acknowledgments of the SARS-COV-2 genomes are also given in the Supporting Material.

\section*{Acknowledgment}
This work was supported in part by NIH grant  GM126189, NSF Grants DMS-1721024,  DMS-1761320, and IIS1900473,  Michigan Economic Development Corporation,  Bristol-Myers Squibb, and Pfizer. The authors thank The IBM TJ Watson Research Center, The COVID-19 High Performance Computing Consortium, and  NVIDIA for computational assistance.


\end{document}